%% file: main.tex
\documentclass[sigconf]{acmart}

\usepackage{amsmath}
\usepackage{braket}
\usepackage{graphicx}
\usepackage{subcaption}

\AtBeginDocument{%
  \providecommand\BibTeX{{%
    \normalfont B\kern-0.5em{\scshape i\kern-0.25em b}\kern-0.8em\TeX}}}

\copyrightyear{2019} 
\acmYear{2019} 
\acmConference[MICRO-52]{The 52nd Annual IEEE/ACM International Symposium on Microarchitecture}{October 12--16, 2019}{Columbus, OH, USA}
\acmBooktitle{The 52nd Annual IEEE/ACM International Symposium on Microarchitecture (MICRO-52), October 12--16, 2019, Columbus, OH, USA}
\acmPrice{15.00}
\acmDOI{10.1145/3352460.3358313}
\acmISBN{978-1-4503-6938-1/19/10}

\setcopyright{acmcopyright}

\begin{document}

\title{Partial Compilation of Variational Algorithms for Noisy Intermediate-Scale Quantum Machines}

\author{Pranav Gokhale}
\authornote{Corresponding author: pranavgokhale@uchicago.edu}
\affiliation{\institution{University of Chicago}}
\orcid{0000-0003-1946-4537}

\author{Yongshan Ding}
\affiliation{\institution{University of Chicago}}

\author{Thomas Propson}
\affiliation{\institution{University of Chicago}}

\author{Christopher Winkler}
\affiliation{\institution{University of Chicago}}

\author{Nelson Leung}
\affiliation{\institution{University of Chicago}}

\author{Yunong Shi}
\affiliation{\institution{University of Chicago}}

\author{David I. Schuster}
\affiliation{\institution{University of Chicago}}

\author{Henry Hoffmann}
\affiliation{\institution{University of Chicago}}

\author{Frederic T. Chong}
\affiliation{\institution{University of Chicago}}

\renewcommand{\shortauthors}{Gokhale et al.}

\input{txt/0abstract.tex}



\maketitle

\input{txt/1introduction.tex}
\input{txt/2background.tex}
\input{txt/2related_work.tex}
\input{txt/3variational_algorithm_benchmarks.tex}
\input{txt/5full_quantum_optimal_control.tex}
\input{txt/6strict_partial_compilation.tex}
\input{txt/7flexible_partial_compilation.tex}
\input{txt/8results.tex}
\input{txt/9conclusion.tex}

\begin{acks}
This work is funded in part by EPiQC, an NSF Expedition in Computing, under grant CCF-1730449 and in part by STAQ, under grant NSF Phy-1818914. Pranav Gokhale is
supported by the Department of Defense (DoD) through the
National Defense Science \& Engineering Graduate Fellowship (NDSEG) Program. Additional funding for Henry Hoffmann comes from the DARPA BRASS program and a DoE Early Career Award. This work was completed in part with resources provided by the University of Chicago Research Computing Center.
\end{acks}

\bibliographystyle{ACM-Reference-Format}
\bibliography{refs}

\appendix
\input{txt/4experimental_setup.tex}

\end{document}

%% file: txt/0abstract.tex
\begin{abstract}
Quantum computing is on the cusp of reality with Noisy Intermediate-Scale Quantum (NISQ) machines currently under development and testing.  Some of the most promising algorithms for these machines are \textit{variational} algorithms that employ classical optimization coupled with quantum hardware to evaluate the quality of each candidate solution. Recent work used GRadient Descent Pulse Engineering (GRAPE) to translate quantum programs into highly optimized machine control pulses, resulting in a significant reduction in the execution time of programs.  This is critical, as quantum machines can barely support the execution of short programs before failing.

However, GRAPE suffers from high compilation latency, which is untenable in variational algorithms since compilation is interleaved with computation. We propose two strategies for \textit{partial compilation}, exploiting the structure of variational circuits to pre-compile optimal pulses for specific blocks of gates. Our results indicate significant pulse speedups ranging from 1.5x-3x in typical benchmarks, with only a small fraction of the compilation latency of GRAPE.\end{abstract}

%% file: txt/1introduction.tex
\section{Introduction}
In the Noisy Intermediate-Scale Quantum (NISQ) era, we expect to operate hardware with hundreds or thousands of quantum bits (qubits), acted on by imperfect gates \cite{NISQ}. Moreover, connectivity in these NISQ machines will be sparse and qubits will have modest lifetimes. Given these limitations, NISQ era machines will not be able to execute large-scale quantum algorithms like Shor Factoring \cite{Shor} and Grover Search \cite{Grover}, which rely on error correction that requires millions of qubits \cite{MagicStateOverhead, ErrorCorrectionOverhead}.

However, recently, \textit{variational algorithms} have been introduced that are well matched to NISQ machines. This new class of algorithms has a wide range of applications such as molecular ground state estimation \cite{VQE}, MAXCUT approximation \cite{QAOA}, and prime factorization \cite{VQF}. The two defining features of a variational algorithm are that:
\begin{enumerate}
    \item the algorithm complies with the constraints of NISQ hardware. Thus, the circuit for a variational algorithm should have modest requirements in qubit count (circuit width) and runtime (circuit depth / critical path).
    \item the quantum circuit for the algorithm is parametrized by a list of angles. These parameters are optimized by a classical optimizer over the course of many iterations. For this reason, variational algorithms are also termed as hybrid quantum-classical algorithms \cite{NISQ}. Typically, a classical optimizer that is robust to small amounts of noise (e.g. Nelder-Mead) is chosen \cite{VQE, McClean_2016}.
\end{enumerate}

Standard non-variational quantum algorithms are fully specified at compile time and therefore can be fully optimized by static compilation tools as in previous work \cite{ScaffCC, QuantumRotations}. By contrast, each iteration of a variational algorithm depends on the results of the previous iteration--hence, compilation must be interleaved through the computation. As even small instances of variational algorithms will require thousands of iterations \cite{Kandala}, the compilation latency for each iteration therefore becomes a serious limitation. This feature of variational algorithms is a significant departure from previous non-variational quantum algorithms.

To cope with this limitation on compilation latency, past work on variational algorithms has performed compilation under the standard gate-based model. This methodology has the advantage of extremely fast compilation--a lookup table maps each gate to a sequence of machine-level control pulses so that compilation simply amounts to concatenating the pulses corresponding to each gate. We note that this compilation procedure is a conservative picture of experimental approaches to gate-based compilation. In practice, parametrized gates may be handled by a step-function lookup table that depends on the run-time parameters, with the aim of reducing errors, as demonstrated in \cite{WaterVQE, DigitizedAdiabaticQC, PhysRevA.96.022330}.

The gate-based compilation model is known to fall short of the GRadient Ascent Pulse Engineering (GRAPE) \cite{Glaser2015, GRAPE} compilation technique, which compiles directly to the level of the machine-level control pulses that a quantum computer actually executes. In past work \cite{YunongPaper, NelsonPaper, MohamedPaper}, GRAPE has been used to achieve 2-5x pulse speeedups over gate-based compilation for a range of quantum algorithms. Since fidelity decreases exponentially in time, with respect to the extremely short lifetimes of qubits, reducing the pulse duration is critical to ensuring that a computation completes before being completely scrambled by quantum decoherence effects. Thus, 2-5x pulse speedups translate to an even bigger advantage in the success probability of a quantum circuit.

However, GRAPE-based compilation has a substantial cost: compilation time. Running GRAPE control on a circuit with just four qubits takes several minutes. For representative four qubit circuits, we observed compile times ranging from 10 minutes to 1 hour, even with state-of-the-art hardware and GPU acceleration. This would amount to several weeks or months of total compilation latency over the course of thousands of iterations (and millions of iterations will be needed for larger problems). By contrast, typical pulse times for quantum circuits are on the order of microseconds, so the compilation latency imposed by GRAPE is untenable. Thus, GRAPE-based compilation is not practical out-of-the-box for variational algorithms.

In this paper, we introduce \textit{partial compilation}, a strategy that approaches the pulse duration speedup of GRAPE, but with a manageable overhead in compilation latency. With this powerful new compiler capability, \textbf{we enable the architectural choice of pulse-level instructions}, which supports more complex near-term applications through lower latencies and thus much lower error rates. This architectural choice would be infeasible without our compiler support. Our specific contributions include:
\begin{itemize}
    \item Demonstration of the advantage of GRAPE over gate based compilation for variational algorithms
    \item Strict partial compilation, a strategy that pre-computes optimal pulses for parametrization-independent blocks of gates. This strategy is strictly better than gate-based compilation--it achieves a significant pulse speedup (approaching GRAPE results), with no overhead in compilation latency.
    \item Flexible partial compilation, a strategy that performs as well as full GRAPE, but with a dramatic speedup in compilation latency via precomputed hyperparameter optimization.
\end{itemize}

The rest of this paper is organized as follows. Section~\ref{sec:background} gives prerequisite background on quantum computation and Section~\ref{sec:related_work} describes related work from prior research. Section~\ref{sec:variational_algorithm_benchmarks} describes characteristics of our benchmark variational algorithms, with particular attention to the structural properties that our compilation strategies exploit. Section~\ref{sec:full_qoc} explains the GRAPE compilation methodology. Sections \ref{sec:strict_partial} and \ref{sec:flexible_partial} explain our partial compilation strategies and Section~\ref{sec:results} discusses our results. We conclude in Section~\ref{sec:conclusion} and propose future work in Section~\ref{sec:future_work}. Appendix~\ref{sec:experimental_setup} presents the system Hamiltonian that we consider in GRAPE.

%% file: txt/2background.tex
\section{Background on Quantum Computation} \label{sec:background}

\subsection{Qubits}

The fundamental unit of quantum computation is a quantum bit, or qubit. A qubit has two basis states, which are represented by \textit{state vectors} denoted
$$\ket{0} = \begin{pmatrix} 1 \\  0 \end{pmatrix} \text{ and } \ket{1} = \begin{pmatrix} 0 \\ 1 \end{pmatrix}.$$

Unlike a classical bit, the state of a qubit can be in a \textit{superposition} of both $\ket{0}$ and $\ket{1}$. In particular, the space of valid qubit states are $\alpha \ket{0} + \beta \ket{1}$, normalized such that $|\alpha|^2 + |\beta|^2 = 1$. When a qubit is measured, its quantum state \textit{collapses} and either $\ket{0}$ or $\ket{1}$ is measured, with probabilities $|\alpha|^2$ and $|\beta|^2$ respectively.

A two-qubit system has four basis states:
$$\ket{00} = \begin{pmatrix} 1 \\ 0 \\ 0 \\  0 \end{pmatrix} \text{, } \ket{01} = \begin{pmatrix} 0 \\ 1 \\ 0 \\ 0 \end{pmatrix}, \ket{10} = \begin{pmatrix} 0 \\ 0 \\ 1 \\  0 \end{pmatrix} \text{, and } \ket{11} = \begin{pmatrix} 0 \\ 0 \\ 0 \\ 1 \end{pmatrix}$$
and any two-qubit state can be expressed as the superposition $\alpha \ket{00} + \beta \ket{01} + \gamma \ket{10} + \delta \ket{11}$ (normalized so that $|\alpha| + |\beta|^2 + |\gamma|^2 + |\delta|^2 = 1$). More generally, an $N$-qubit system has $2^N$ basis states. Therefore, $2^N$ numbers, called amplitudes, are needed to describe the state of a general $N$-qubit system. This exponential scaling gives rise to both the difficulty of classically simulating a quantum system, as well as the potential for quantum computers to exponentially outperform classical computers in certain applications.

\subsection{Quantum Gates}
A quantum algorithm is described in terms of a quantum circuit, which is a sequence of 1- and 2- input quantum gates. Every quantum gate is represented by a square matrix, and the action of a gate is to left-multiply a state vector by the gate's matrix. Because quantum states are normalized by measurement probabilities, these matrices must preserve $l^2$-norms. This corresponding set of matrices are \textit{unitary} (orthogonal) matrices. The unitary matrices for two important single-qubit gates are:
$$R_x(\theta) = \begin{pmatrix} i \cos{\frac{\theta}{2}} & \sin{\frac{\theta}{2}} \\ \sin{\frac{\theta}{2}} & i \cos{\frac{\theta}{2}} \end{pmatrix} \text{ and } R_z(\phi) = \begin{pmatrix} 1 & 0 \\ 0 & e^{i \phi} \end{pmatrix}$$

At $\theta = \pi$, the $R_x(\pi)$ gate has matrix $\bigl( \begin{smallmatrix} 0 & 1 \\ 1 & 0 \end{smallmatrix} \bigl)$, which acts as a NOT gate: left-multiplying by it swaps between the $\ket{0}$ and $\ket{1}$ states. This bit-flip gate is termed the $X$ gate.

Similarly, at $\phi = \pi$, the $R_z(\pi)$ gate has matrix $\bigl( \begin{smallmatrix} 1 & 0 \\ 0 & -1 \end{smallmatrix} \bigl)$, which applies a $-1$ multiplier to the amplitude of $\ket{1}$; this type of gate is unique to the quantum setting, where amplitudes can be negative (or complex). This `phase'-flip gate is termed the $Z$ gate.

An important 2-input quantum gate is
$$\text{CX} = \begin{pmatrix} 1 & 0 & 0 & 0 \\ 0 & 1 & 0 & 0 \\ 0 & 0 & 0 & 1 \\ 0 & 0 & 1 & 0 \end{pmatrix}$$
The CX gate, often referred to as the CNOT or Controlled-NOT gate, applies an action that is controlled on the first input. If the first input is $\ket{0}$, then the CX gate has no effect. If the first input is $\ket{1}$, then it applies an $X = R_x(\pi)$ to the second qubit.

The CX gate is an \textit{entangling gate}, meaning that its effect cannot be decomposed into independent gates acting separately on the two qubits. An important result in quantum computation states that the set of all one qubit gates, plus a single entangling gate, is sufficient for universality \cite{Nielsen}. Since the $R_x(\theta)$ and $R_z(\phi)$ gates span the set of all one qubit gates, we see that, $\{R_x(\theta), R_z(\phi), CX\}$ is a universal gate set.

In practice, we seek to implement a quantum algorithm using the most efficient quantum circuit possible, with efficiency defined in terms of circuit width (number of qubits) and depth (length of critical path, or runtime of the circuit). Accordingly, quantum circuits are optimized by repeatedly applying gate identities that reduce the resources consumed by the circuit. All circuits that are presented in this paper were optimized using IBM Qiskit's Transpiler, which applies a variety of circuit identities--for example, aggressive cancellation of CX gates and `Hadamard' gates. We also augmented the IBM optimizer with our own compiler pass for merging rotation gates--e.g. $R_x(\alpha)$ followed by $R_x(\beta)$ merges into $R_x(\alpha + \beta)$--which we found to further reduce circuit sizes.

\subsection{Gate-Based Compilation}
At the lowest level of hardware, quantum computers are controlled by analog pulses. Therefore, quantum compilation must translate from a high level quantum algorithm down to a sequence of control pulses. Once a quantum algorithm has been decomposed into a quantum circuit comprising single- and two- qubit gates, gate-based compilation simply proceeds by concatenating a sequence of pulses corresponding to each gate. In particular, a lookup table maps from each gate in the gate set to a sequence of control pulses that executes that gate. Table~\ref{tab:gate_times} indicates the total pulse duration for each gate in the compilation basis gate set. These pulse durations are based on the gmon-qubit \cite{Gmon} quantum system described in Appendix~\ref{sec:experimental_setup}.

\begin{table}[t]
\renewcommand{\arraystretch}{1.3}
\centering
\begin{tabular}{|c|c|c|c|c|c|c}
\hline      
Gate & $R_z(\phi)$ & $R_x(\theta)$ & H & CX & SWAP \\
\hline
Time (ns) & 0.4 & 2.5 & 1.4 & 3.8 & 7.4 \\
\hline
\end{tabular}
\newline
\vspace{0.15cm}
\caption{Library of the compiler's gate set and corresponding pulse durations (in nanoseconds) for each gate. The runtimes of circuits under gate-based compilation are indexed to these pulse durations.}
\label{tab:gate_times}
\end{table}

As previously noted, the $\{R_x(\theta), R_z(\phi), \text{CX}\}$ gate set alone is sufficient for universality, so in principle the H and SWAP gates could be removed from the compilation basis gate set. However, we include the generated pulses (using GRAPE as described below) for these gates in our compilation set, because quantum assembly languages typically include them in their basis set \cite{OpenPulse, ProjectQ, QSharp, Quil, Quipper, Scaffold}. 

The advantage of the gate-based approach is its short pulse compilation time, as the lookup and concatenation of pulses can be accomplished almost instantaneously. However, it prevents the optimization of pulses from happening across the gates, because there might exists a global pulse for the entire circuit that is shorter and more accurate than the concatenated one. The quality of the concatenated pulse relies heavily on an efficient gate decomposition of the quantum algorithm. 

\subsection{GRAPE}
GRadient Pulse Engineering (GRAPE) is a strategy for compilation that numerically finds the best control pulses needed to execute a quantum circuit or subcircuit by following a gradient descent procedure \cite{GRAPE, GRAPE2}. We use the Tensorflow implementation of GRAPE described in \cite{NelsonPaper}. In contrast to the gate based approach, GRAPE does not have the limitation incurred by the gate decomposition. Instead, it directly searches for the optimal control pulse for the input circuit as a whole. Our full GRAPE procedure is described further detail in Section~\ref{sec:full_qoc}.

%% file: txt/2related_work.tex
\section{Related Work} \label{sec:related_work}

Past publications of variational algorithm implementations have relied on gate-based compilation, using parametrized gates such as $R_x(\theta)$ and $R_z(\phi)$. Existing quantum languages offer support for such parametrized gates \cite{Scaffold, OpenPulse, ProjectQ, LIQUI, Quil, Quipper}. In most languages, the angles must be declared at compile time--thus at every iteration of a variational algorithm, a new circuit is compiled based on the new parametrization. Rigetti's Quil \cite{Quil} language goes a step further by supporting runtime resolution of the parameters in parameters gates, which allows dynamic implementations of variational algorithms. However, as acknowledged in the Quil specifications, this approach hampers circuit optimization, because the actual parameters are not known until runtime.

While this paper treats gate-based compilation as a simple lookup table between gates and pulses, experimental implementations have already moved directionally towards GRAPE-style, because pulse sequences can depend on the input angles in a complicated fashion. For example, in \cite{DigitizedAdiabaticQC}, a parametrized $U(\phi)$ gate has five different pulse sequence decompositions, each corresponding to $\phi$ in ranges set by the breakpoints $[-\pi, 2.25, -0.25, 0.25, 2.25, \pi]$. \cite{WaterVQE} and \cite{PhysRevA.96.022330} have similar step-function gate-to-pulse translation.

The growing overhead of compilation latency has been recognized, and recent work has proposed the development of specialized FPGAs for the compilation of variational algorithms \cite{Moll_2018}. More broadly, we note that pulse level control is at the cusp of industry adoption. An open specification for pulse-level control, OpenPulse, was standardized recently \cite{OpenPulse}, and IBM plans to introduce an API for pulse level control in 2019 \cite{QiskitRoadmap}. Pulse access to quantum machines will open the door to experimental realizations of GRAPE, including for variational algorithms as proposed in this paper.


%% file: txt/3variational_algorithm_benchmarks.tex
\section{Variational Benchmarks} \label{sec:variational_algorithm_benchmarks}
\begin{figure}[t]
    \centering
    \includegraphics[width=0.85\linewidth]{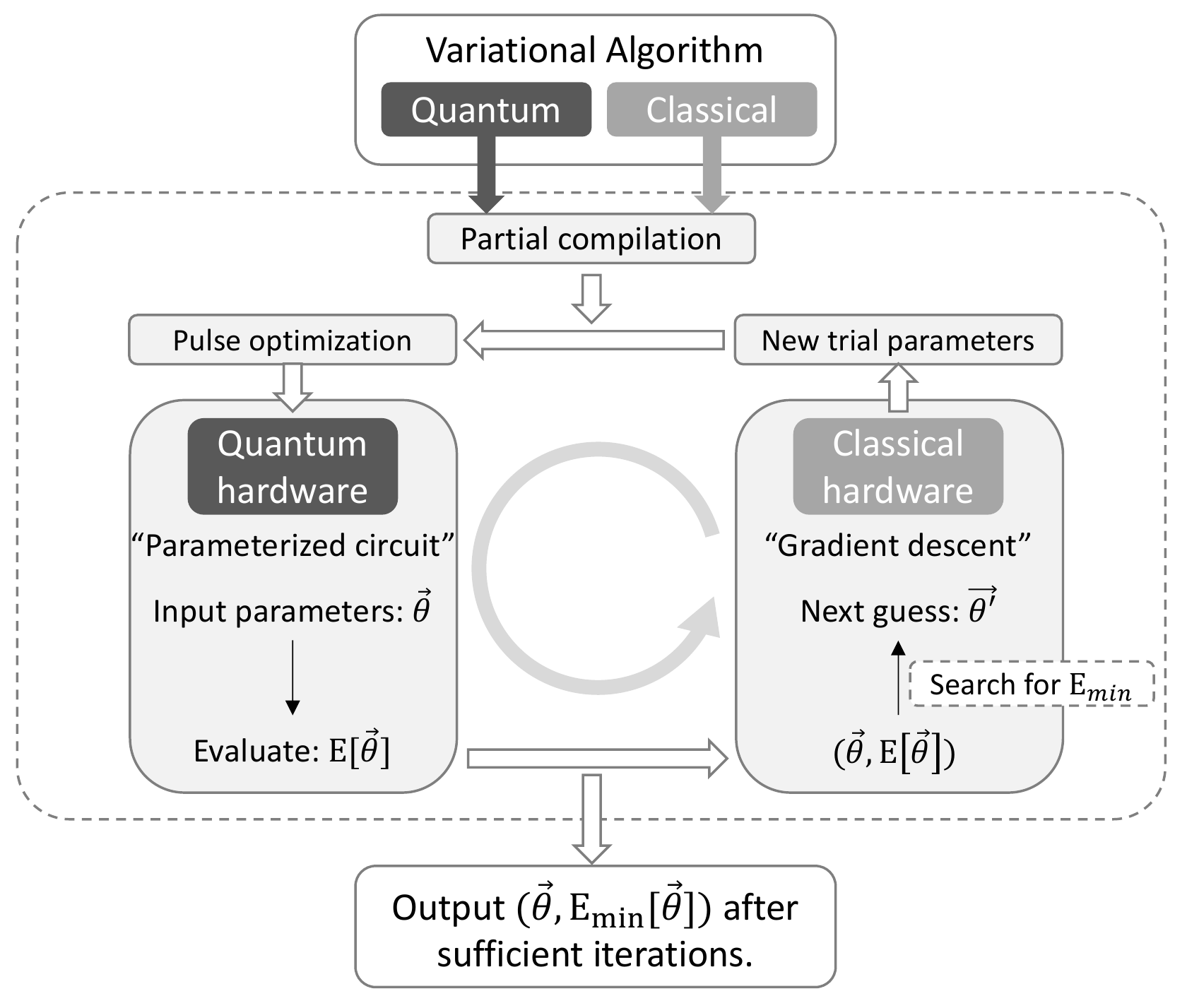}
    \caption{Illustration of a variational quantum algorithm that alternates between a quantum circuit and a classical optimizer. In this process, the quantum circuit (parameterized by $\vec{\theta}$) evaluates some cost function $E[\vec{\theta}]$, and the classical optimizer gradient descends for the next set of parameters.}
    \label{fig:var_algo}
\end{figure}

Variational quantum algorithms are important in the near-term because they comply with the constraints of NISQ hardware. In particular, variational algorithms have innate error resilience, due to the hybrid alternation with a noise-robust classical optimizer \cite{VQE, McClean_2016}. Every iteration of a variational algorithm is parameterized by a list of angles. In general, the parameter space explored by a variational algorithm is not known a priori--the classical optimizer picks the next iteration's parameters based on the results of the previous iterations. Consequently, the compilation for each iteration is interleaved with the actual computation. A schematic of this process is illustrated in Figure \ref{fig:var_algo}.

There are two variational quantum algorithms: Variational Quantum Eigensolver and Quantum Approximate Optimization Algorithm. We discuss both below.
\subsection{Variational Quantum Eigensolver}
The Variational Quantum Eigensolver (VQE) is used to find the ground state energy of a molecule, a task that is exponentially difficult in general for a classical computer, but is believed to be efficiently solvable by a quantum computer \cite{Quantum_Computational_Chemistry}. Estimating the molecular ground state has important applications to chemistry such as determining reaction rates \cite{Eyring} and molecular geometry \cite{1206.2247}. A conventional quantum algorithm for solving this problem is called the Quantum Phase Estimation (QPE) algorithm \cite{lloyd1996universal}. However, for a target precision $\epsilon$, QPE requires a quantum circuit with depth $O(1/\epsilon)$, whereas $VQE$ algorithm requires $O(1/\epsilon^2)$ iterations of depth-$O(1)$ circuits\cite{wang2018generalised}. The latter assumes a much more relaxed fidelity requirement on the qubits and gate operations, because the higher the circuit depth, the more likely the circuit experiences an error at the end. At a high level, VQE can be conceptualized as a guess-check-repeat algorithm. The check stage involves the preparation of a quantum state corresponding to the guess. This preparation stage is done in polynomial time on a quantum computer, but would incur exponential cost (owing to the $2^N$ state vector scaling) in general on a classical computer. This contrast gives rise to a potential quantum speedup for VQE.

The quantum circuit corresponding to the guess is termed an ansatz. While many ansatz choices are possible, Unitary Coupled Cluster Single-Double (UCCSD), an ansatz motivated by principles of quantum chemistry, is considered the gold standard \cite{UCCSD_Strategies, RevModPhys.79.291}. The UCCSD ansatz is also promising because it could circumvent the barren plateaus issue that affects many other ansatzes \cite{Quantum_Computational_Chemistry}.

We benchmark the UCCSD ansatz for five molecules: H\textsubscript{2}, LiH, BeH\textsubscript{2}, NaH, H\textsubscript{2}O. These molecules span the state of the art for experimental implementations of VQE: H\textsubscript{2}O is the largest molecule addressed by VQE \cite{WaterVQE} to date. We generated our UCCSD ansatz circuits using the IBM Qiskit implementation described in \cite{IBM_UCCSD} as well as the PySCF Python package \cite{PySCF} to manage molecular data.

Both the the circuit depth and number of ansatz parameters in UCCSD scale as $O(N^4)$ in the circuit width \cite{Babbush2015}. Table~\ref{tab:vqe} specifies the exact circuit width, number of variational parameters, and gate-based runtime (circuit depth) for each of the benchmarks. The reported gate-based runtimes are indexed to the pulse durations of each gate reported in Table~\ref{tab:gate_times}. Each circuit was optimized using IBM Qiskit's circuit optimizer pass system, Qiskit's circuit mapper (to conform to nearest neighbor connectivity), and a custom compiler pass to merge neighboring rotation gates on the same axis. We also exploit parallelism to simultaneously schedule as many gates as posisble; the reported gate-based runtimes are for the critical path through the parallelized circuit. These circuit optimizations form a fair baseline for the best circuit runtimes achievable by gate based compilation. Our full circuit optimization code, along with the results of optimization applied to our benchmarks, is available on our Github repository \cite{GithubRepo}.

\renewcommand{\arraystretch}{1.4} 
\begin{table}[]
\small
\centering
\begin{tabular}{|c|c|c|c|ll}
\cline{1-4}
Molecule              &  Width (\# of Qubits) & \# of Params    & \scriptsize{Gate-Based Runtime} &  \\ \cline{1-4}
H\textsubscript{2}    & 2                   & 3                       & 35 ns      &  \\ \cline{1-4}
LiH                   & 4                   & 8                       & 872 ns     &  \\ \cline{1-4}
BeH\textsubscript{2} & 6                   & 26                      & 5308 ns    &  \\ \cline{1-4}
NaH                   & 8                   & 24                      & 5490 ns    &  \\ \cline{1-4}
H\textsubscript{2}O   & 10                  & 92                      & 33842 ns   &  \\ \cline{1-4}
\end{tabular}
\caption{Benchmarked circuits for VQE, using the UCCSD ansatz. Each circuit was optimized, parallel-scheduled, mapped using IBM Qiskit's tools, augmented by an additional optimization pass we wrote to merge consecutive rotation gates. The Gate-Based Runtime is indexed to the pulse durations for each gate reported in Table~\ref{tab:gate_times}.}
\label{tab:vqe}
\end{table}

\subsection{QAOA}
Quantum Approximate Optimization Algorithm (QAOA) is an algorithm for generating approximate solutions to problems that are hard to solve exactly. At an intuitive level, QAOA can be understood as an alternating pattern of Mixing and Cost-Optimization steps. At each Mixing step, QAOA applies diffusion so that every possible state is explored in quantum superposition. At each Cost-Optimization step, a bias is applied to boost the magnitudes of quantum states that minimize a cost function. Thereafter, measuring can yield an approximate solution close to optimal with high probability. The number of alternating Mixing and Cost-Optimization rounds is known as $p$. Even for small $p$, QAOA has competitive results against classical approximation algorithms. For example, at $p = 1$, QAOA applied to the NP-hard MAXCUT problem yields a cut of size at least 69\% of the optimal cut size \cite{QAOA}. At $p=5$, simulations have demonstrated that QAOA achieves mean parity with the best-known classical algorithm, Goemans-Williamson, for 10 node graphs \cite{Rigetti_QAOA_Simulation}. For larger $p$, QAOA is expected to outperform classical approximation algorithms even for worst-case bounds, although theoretical guarantees have not been established yet. QAOA is of particular interest in the near term because recent work has shown that it is computationally universal \cite{QAOA_Universality}. Moreover, QAOA has shown experimental resilience to noise \cite{Rigetti_QAOA_Experiment}. For these reasons, QAOA is a leading candidate for quantum supremacy \cite{QAOA_Supremacy}, the solution of a classically-infeasible problem using a quantum computer.

Similarly to VQE, QAOA is a guess-check-repeat algorithm. In the case of QAOA, the guesses correspond to ``Mixing magnitude during iteration $1 \leq i \leq p$" and ``Cost-Optimization magnitude during iteration $1 \leq i \leq p$". Hence, the number of parameters in a QAOA circuit is $2p$: one scalar for Mixing magnitude and one for Cost-Optimization magnitude, for each of the $p$ rounds.

We benchmark QAOA for $N =$ 6 and 8 node graphs, with the number of QAOA rounds $p$ spanning from 1 to 8. For each $(N, p)$ pair, we benchmark for two types of random graphs: 3-regular (each node is connected to three neighbors) and Erdos-Renyi (each possible edge is included with 50\% probability). This yields $2 \times 8 \times 2 = 32$ benchmarks circuits for QAOA. The gate-based runtimes for each of these benchmarks are reported in Table~\ref{tab:qaoa}. As with the VQE benchmarks, the runtimes are computed after circuit mapping and optimizations, to form a fair baseline.

\renewcommand{\arraystretch}{1.4} 
\begin{table}[]
\small
\centering
\begin{tabular}{|c|c|c|c|c|ll}
\cline{2-5}
\multicolumn{1}{c|}{} &  \multicolumn{2}{c|}{$N=6$}  & \multicolumn{2}{c|}{$N=8$} & \\ \cline{2-5}
\multicolumn{1}{c|}{} &  3-Regular & Erdos-Renyi  & 3-Regular & Erdos-Renyi & \\ \cline{1-5}
$p=1$  &  113 ns & 84 ns  & 163 ns & 157 ns & \\ \cline{1-5}
$p=2$  &  199 ns & 151 ns  & 365 ns & 297 ns & \\ \cline{1-5}
$p=3$  &  277 ns & 223 ns  & 530 ns & 443 ns & \\ \cline{1-5}
$p=4$  &  356 ns & 296 ns  & 695 ns & 596 ns & \\ \cline{1-5}
$p=5$  &  434 ns & 368 ns  & 860 ns & 750 ns & \\ \cline{1-5}
$p=6$  &  512 ns & 440 ns  & 1025 ns & 903 ns & \\ \cline{1-5}
$p=7$  &  590 ns & 512 ns  & 1191 ns & 1056 ns & \\ \cline{1-5}
$p=8$  &  668 ns & 584 ns  & 1356 ns & 1209 ns & \\ \cline{1-5}

\end{tabular}
\caption{Gate-based runtimes for our 32 benchmark QAOA MAXCUT circuits. Our benchmarks consider two types of random graphs: 3-Regular and Erdos-Renyi. We consider both 6 and 8 node graphs--the number of qubits in the circuit is the same as the number of nodes in the graph. We benchmarked over $p$, the number of repetitions of the basic QAOA block, ranging from 1 to 8, which represents a range of $p$ that is of both theoretical and practical interest \cite{Rigetti_QAOA_Simulation}. As in Table~\ref{tab:vqe}, the gate-based runtimes are based on the gate times in Table~\ref{tab:gate_times}, after each circuit has been optimized, parallel-scheduled, and mapped.}
\label{tab:qaoa}
\end{table}


%% file: txt/5full_quantum_optimal_control.tex
\section{GRAPE Compilation} \label{sec:full_qoc}
In this section, we describe GRAPE (GRadient Ascent Pulse Engineering), a compilation technique that aims to produce the optimal possible sequence of analog control pulses needed to realize the unitary matrix transformation for a targeted quantum circuit. At an abstract level, GRAPE simply treats the underlying quantum computer as a black box. The black box accepts time-discretized control pulses as input and outputs the unitary matrix of the transformation that is realized by the input control pulses. GRAPE performs gradient descent over the space of possible control pulses to search for the optimal sequence of input signals that achieve the targeted unitary matrix up to a specified fidelity. We used the Tensorflow-based implementation of GRAPE described in \cite{NelsonPaper}, which has demonstrated good performance. The gradients are computed analytically and backpropogated with automatic differentiation.

In this paper, we define the optimal sequence of control pulse as the one of shortest duration--thus, we seek to speed up the pulse time with respect to gate-based compilation. Reducing the pulse time is important in quantum computation because qubits have short lifetimes due to quantum decoherence effects. The decoherence error increases exponential with time, so the effect of a pulse time speedup enters the power of an exponential term. We focus on this error metric because it is one of the dominant error terms for superconducting qubits and it is well understood. However, in principle, GRAPE can be used to control other sources of error such as gate errors, State Preparation and Measurement (SPAM) errors, and qubit crosstalk, as demonstrated in past work \cite{MohamedPaper, QOC_Crosstalk, PhysRevA.91.052315}.

\subsection{Speedup Sources}

Because GRAPE translates directly from a unitary matrix to hardware-level control pulses--without the overhead of an intermediate set of quantum gates--it achieves more optimized control pulses than gate-based compilation does. In particular, we observed significant pulse speedups from GRAPE due to the following factors:
\begin{itemize}
    \item \textbf{ISA alignment}. Gate based compilation incurs a significant overhead because the set of basis gates will not be \textit{directly} implementable on a target machine. For example, while quantum circuits are typically compiled down to CX (CNOT) gates as the default two-qubit instructions, actual quantum computers implement a wide range of native two-qubit operations such as the MS gate or the iSWAP gate. Compiling gates to pulses incurs a significant overhead from this ISA misalignment.
    \item \textbf{Fractional gates}. A unique feature of quantum computing is that all operations can be fractionally performed--for example, $\text{CX}^{1/2}$ is a valid quantum gate, as is $\text{CX}^p$ more generally for any power. Often, a fractional application of a basis gate is sufficient to execute a larger quantum operation. The fixed basis set of gate based compilation misses these optimizations, whereas GRAPE works in a continuous basis and realizes fractional gates when beneficial.
    \item \textbf{Control Field Asymmetries}. While gate based compilation puts $R_x$ and $R_z$ gates on an equal footing, at a physical level, there is often a significant asymmetry between the speed and reliability of these operations. As described in A, we model a representative quantum system in which $Z$-axis qubit rotations are 15 times faster than $X$-axis qubit rotations. GRAPE's search for the shortest pulse realization will therefore leverage this asymmetry, preferring $Z$ rotations when possible. For example, the $H$ gate is typically implemented by the $R_x(\frac{-\pi}{2}) R_z(\frac{-\pi}{2}) R_x(\frac{-\pi}{2})$ pulse sequence, which involves two $X$-axis rotations and one $Z$-axis rotation. We observe that our GRAPE system instead discovers the equivalent $R_z(\frac{-\pi}{2}) R_x (\frac{-\pi}{2}) R_z(\frac{-\pi}{2})$ pulse sequence, which only requires one $X$-axis rotation and therefore executes significantly faster.
    \item \textbf{Maximal circuit optimization}. Although quantum circuits can be optimized at the gate-level by repeatedly applying a set of circuit identity templates, the set of templates must be finite. Opportunities for optimization between distant gates (both in width and depth) may be overlooked. By contrast, GRAPE subsumes all circuit optimizations by working directly in terms of the unitary matrix of the circuit, as opposed to the gate decomposition.
\end{itemize}

\subsection{Circuit Blocking for GRAPE}
\label{subsec:circuit_blocking}
While GRAPE can achieve significant pulse speedups, it is limited by two factors:
\begin{itemize}
    \item The unitary matrix of the targeted quantum circuit must be specified as input to the GRAPE program. An $N$-qubit circuit has a $2^N \times 2^N$ matrix (due to the exponential state space of an $N$-qubit space), which imposes a bound on the maximum circuit size that GRAPE can handle.
    \item The total convergence time for GRAPE's gradient descent scales exponentially in the size of the target quantum circuit \cite{NelsonPaper}. For example, it typically takes our GRAPE implementation several minutes to find the pulses for a 4 qubit QAOA MAXCUT circuit. Experientially, we also found difficulty consistently finding convergence for deep quantum circuits with $N > 5$ qubits.
\end{itemize}

For this reason, it is necessary to partition large quantum circuits into blocks of manageable width. We blocked into subcircuits of up to 4 qubits, using the aggregation methodology discussed in \cite{YunongPaper}. Specifically, we select maximal subcircuits of 4 qubit width, such that partitioning the subcircuit does not delay the execution of subcircuits. This methodology ensures that full GRAPE is strictly better than gate based compilation--otherwise, subcircuits may induce serialization that underperforms gate based compilation. Details are discussed in Section 4.3 of \cite{YunongPaper}.

\subsection{Binary Search for Minimum Pulse Time}
In prior work \cite{NelsonPaper, YunongPaper}, the pulse length is specified as a static `upper bound' parameter, \texttt{total\_time}. Pulse speedups are then performed by adding a term to the cost function that rewards pulses that realize the targeted unitary matrix in time shorter than \texttt{total\_time}. However, to comply with the automatic differentiation methodology for analytically computing gradients, this cost function term is continuous and rewards \textit{gradual} progress of the pulse towards the target unitary matrix. By contrast, our ultimate goal is to find the \textit{binary} cutoff point specifying the minimal possible time needed to achieve a pulse. Moreover, setting the relative weighting of the speedup term to the fidelity term in the cost function is difficult. Poor choices of weights can either prevent GRAPE from achieving any speedup or realizing the target fidelity.

As proposed by the prior work \cite{NelsonPaper}, our methodology adaptively changes the \texttt{total\_time} by binary searching for the shortest \texttt{total\_time} needed to achieve a target unitary matrix. While this incurs the overhead of running on the more iterations \footnote{Specifically, on the order of $log(M/\Delta t)$ iterations where $M$ is the upper bound on \texttt{total\_time} and $\Delta t$ is the desired precision, which we set to 0.3 ns.}, it is worthwhile because minimizing the pulse time is exponentially critical in terms of reducing errors.

\subsection{GRAPE Compilation for QAOA}
There are a range of theoretical results setting upper bounds on the circuit complexity needed to achieve a particular quantum operation. For example, it is known that 3 CX gates, sandwiched by single-qubit rotations, is sufficient to implement any two qubit operation. These results were recently generalized to the context of quantum optimal control (a generalization of GRAPE) with a proof that any $N$-qubit operation can be achieved in $O(4^N)$ time via optimal control \cite{LloydQOC}.

\begin{figure}
    \includegraphics[width=0.45\textwidth]{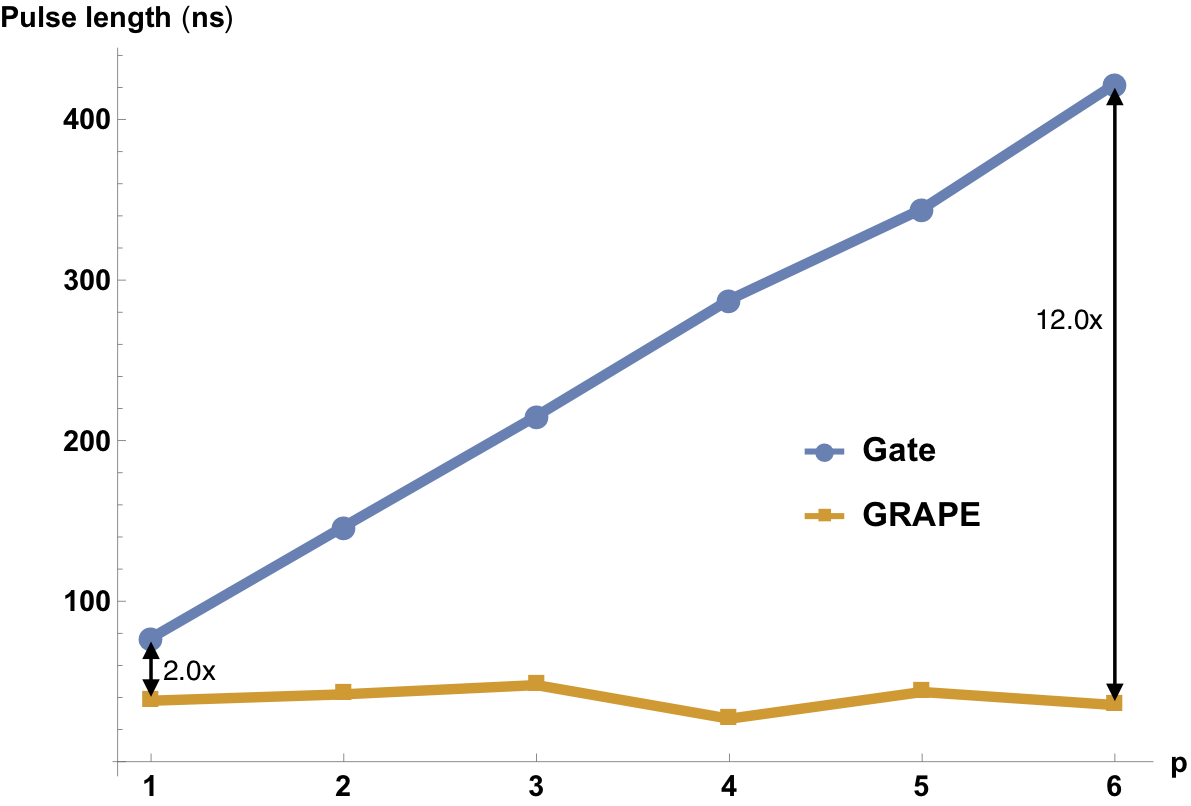}
      \caption{Pulse lengths from gate based compilation and full GRAPE for MAXCUT on the 4-node clique. While the gate based pulse times are simply linear in the number of QAOA rounds $p$, the GRAPE based times asymptote to an upper bound. For each $p$, a random parametrization was set. The ratio varies from 2.0x at $p=1$ to 12.0x at $p=6$.}
      \label{fig:gate_vs_qoc}
\end{figure}

This implies that GRAPE can achieve a significant advantage over gate-based compilation in algorithms like QAOA that have $p$ repeated blocks. While the pulse length from gate-based compilation scales linearly in the $p$, the GRAPE based pulse length is upper bounded by the maximum time it takes to implement any transformation for an $N$-qubit circuit. Figure~\ref{fig:gate_vs_qoc} demonstrates this behavior for QAOA MAXCUT on the 4-node clique problem. While the pulse length from gate based compilation scales linearly in the number of QAOA rounds $p$, it asymptotes below 50 ns for GRAPE based pulse lengths. Thus, the pulse speedup advantage of GRAPE increases with $p$.

As our QAOA benchmarks have circuit widths of 6 and 8 qubits--larger than the 4 qubit blocks we feed to GRAPE--the number of serial blocks will scale linearly with $p$. Therefore, we don't expect to see an unboundedly growing speedup of GRAPE with increasing $p$, but we still expect to see gains within each 4 qubit block.


%% file: txt/6strict_partial_compilation.tex
\begin{figure*}
    \centering
    \begin{subfigure}[b]{0.7\textwidth}
        \includegraphics[width=\textwidth]{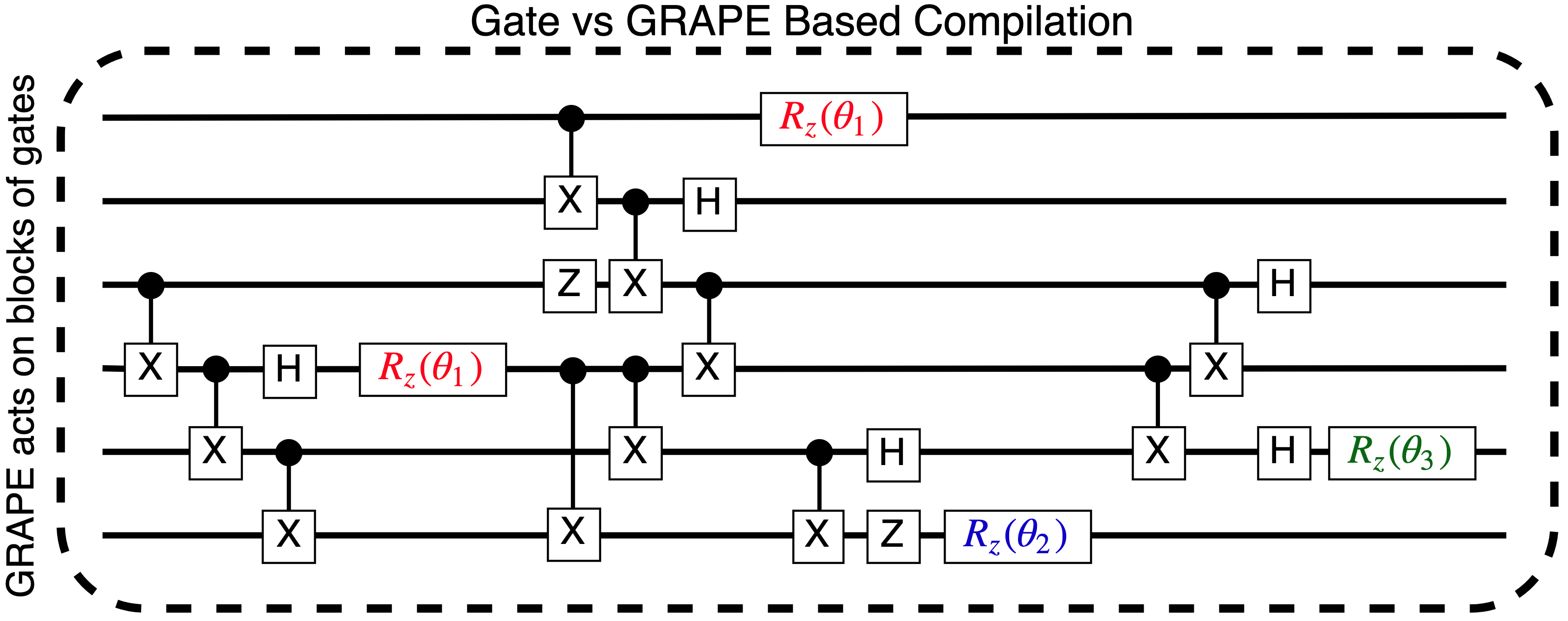}
        \caption{This is a representative variational circuit, decomposed into gates. In gate-based compilation, each gate is translated by a lookup table to analog control pulses. Compilation amounts to simple concatenation of these control pulses. GRAPE (denoted by the dashed line) considers the unitary matrix for the full circuit and performs gradient descent to find the shortest control pulses that realize the circuit. GRAPE achieves significant pulse speedups, but has substantial compilation latency.}
        \label{subfig:typical}
    \end{subfigure}

\par\medskip

    \begin{subfigure}[b]{0.7\textwidth}
        \includegraphics[width=\textwidth]{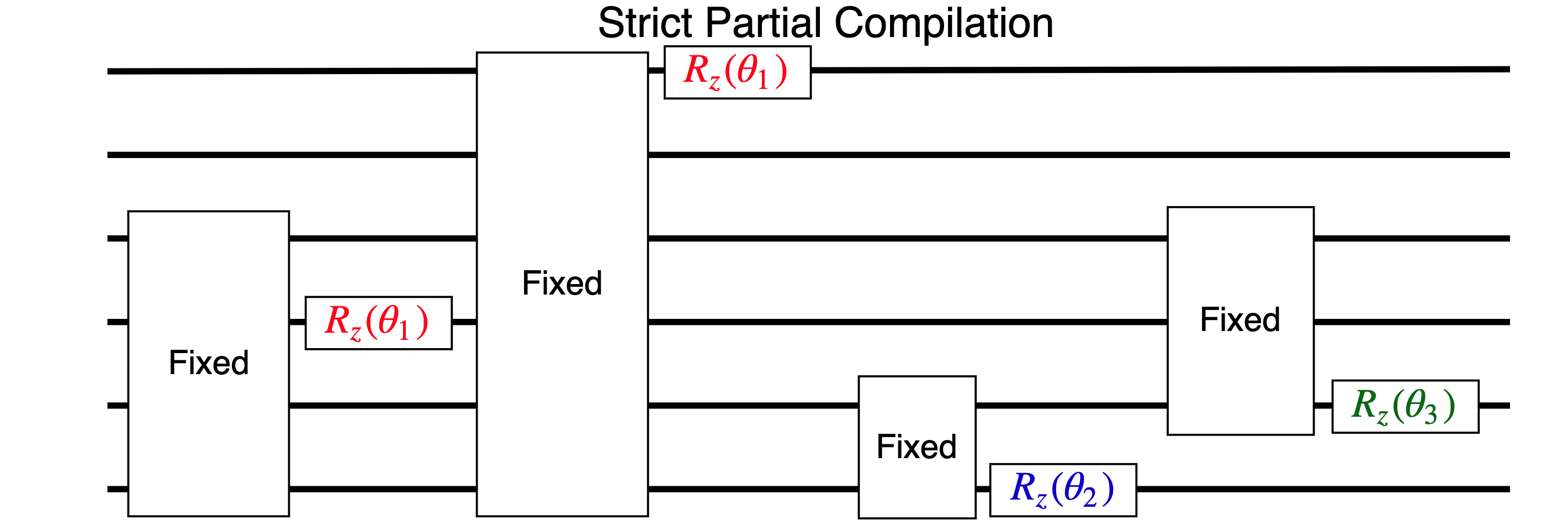}
        \caption{Strict partial compilation blocks the circuit into a strictly alternating sequence of Fixed (parametrization-independent) subcircuits and $R_z(\theta_i)$ gates. Each Fixed subcircuit is precompiled with GRAPE, so that compilation at runtime simply involves concatenating the pulses for each subcircuit.}
        \label{subfig:strict}
    \end{subfigure}

\par\medskip

    \begin{subfigure}[b]{0.7\textwidth}
        \includegraphics[width=\textwidth]{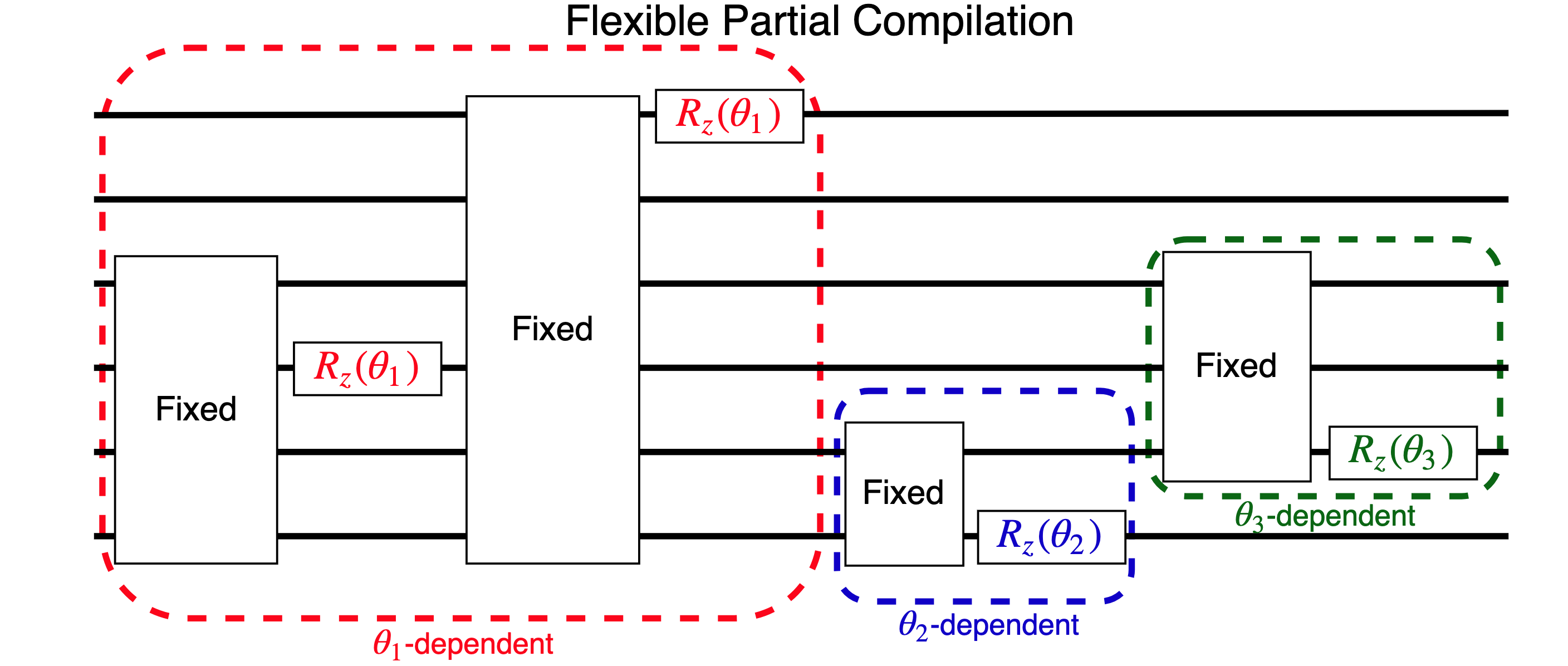}
        \caption{Flexible partial compilation blocks the circuit into subcircuits that depend on exactly one parameter, $\theta_i$. Parameter monotonicity ensures that these subcircuits have significantly longer depth than the Fixed blocks of strict partial compilation. We use hyperparameter optimization to precompute good hyperparameters (learning rate and decay rate) for each subcircuit. When all $\theta_i$ are specified at runtime, we used the tuned hyperparameters to quickly find optimized pulses for each subcircuit.}
        \label{subfig:flexible}
    \end{subfigure}

\par\smallskip

    \caption{Comparison of compilation strategies. Subfigure (a) depicts gate-based and GRAPE- based compilation for a variational circuit. These two compilation approaches represent opposite ends of a spectrum trading off between between compilation latency and control pulse speedup. We introduce two new compilation strategies, strict and flexible partial compilation, that approach the pulse speedup of GRAPE without the large compilation latency. Subfigures (b) and (c) demonstrates strict and flexible partial compilation respectively.}
    \label{fig:compilation_strategies}
\end{figure*}

\section{Strict Partial Compilation} \label{sec:strict_partial}

While full quantum optimal control generates the fastest possible pulse sequence for a target circuit, its compilation latency on the order of several minutes is untenable for variational algorithms, in which compilation is interleaved with computation. In order to approach the pulse speedup of GRAPE without incurring the full cost in compilation latency, it is necessary to exploit the structure of the variational circuits. We term this structural analysis as \textit{partial compilation}, and it is executed as pre-computation step prior to executing the variational algorithm on a quantum computer.

Our first strategy, Strict Partial Compilation, stems from the observation that for typical circuits in variational algorithms, most of the gates are independent of the parametrization. For example, Figure~\ref{subfig:typical} shows an example variational circuit. While the circuit has many gates, only four of them depend on the variational $\theta_i$ parameters. All of the other gates can be blocked into maximal parametrization-independent subcircuits. Figure~\ref{subfig:strict} demonstrates the application of strict partial compilation to the variational circuit from Figure~\ref{subfig:typical}. The sequence of resulting subcircuits is [Fixed, $R_z(\theta_1)$, Fixed, $R_z(\theta_1)$, Fixed, $R_z(\theta_2)$, Fixed, $R_z(\theta_3)$], which exhibits \textit{strict} alternation between `Fixed' subcircuits that don't depend on any $\theta_i$ and $R_z(\theta_i)$ gates that do depend on the parametrization.

After the strict partial compilation blocking is performed, we use full GRAPE to pre-compute the shortest pulse sequence needed to execute each Fixed subcircuit. These static precompiled pulse sequences can be defined as microinstructions in a low-level assembly such as eQASM \cite{eQASM}. Thereafter, at runtime, the pulse sequence for any parametrization can be generated by simply concatenating the pre-computed pulse sequences for Fixed blocks with the control pulses for each parametrization-dependent $R_z(\theta_i)$ gate. Thus, strict partial compilation retains the extremely fast (essentially instant) compilation time of standard gate based compilation. However, since each Fixed block was compiled by GRAPE, the resulting pulse duration is shorter than if the Fixed blocks had been compiled by gate based compilation. Thus, strict partial compilation achieves pulse speedups over gate-based compilation, with no increase in compilation latency.

Full discussion of the results is deferred to Section \ref{sec:results}. A priori, we note that the performance of strict partial compilation is tied to the depth of the Fixed subcircuits. For deeper Fixed subcircuits, GRAPE has more opportunities for optimization and can achieve a greater advantage over gate-based compilation. From inspection of Figure~\ref{subfig:typical}, we see that the depth of Fixed blocks is determined by the frequency of $R_z(\theta_i)$ gates. For our benchmarked VQE-UCCSD circuits, $R_z(\theta_i)$ gates comprise only 5-8\% of the total number of gates, so the Fixed subcircuits have reasonably long depths. For our benchmarked QAOA circuits however, the $R_z(\theta_i)$ gates comprise 15-28\% of the total number of gates, so the Fixed subcircuits have short depths and the potential advantage of strict partial compilation is limited. This motivates us to consider other strategies that more closely match the pulse speedups of full GRAPE.

%% file: txt/7flexible_partial_compilation.tex
\section{Flexible Partial Compilation}
\label{sec:flexible_partial}

As strict partial compilation is bottlenecked by the depth of Fixed subcircuits, we are motivated to consider strategies that create deeper subcircuits. The core idea behind flexible partial compilation is to create subcircuits that are only `slightly' parametrized, in that they depend on at most one of the $\theta_i$ variational parameters. As discussed below, we can perform hyperparameter tuning to ensure that GRAPE finds optimized pulses for single-angle parametrized subcircuits much faster than for general subcircuits.

\subsection{Parameter Monotonicity}

An initial strategy for creating these single-angle parametrized subcircuits would be to merge each consecutive pair of Fixed and $R_z(\theta_i)$ subcircuits into a single subcircuit that only depends on $\theta_i$. However, this strategy would add at most one gate of depth to each subcircuit, which would not lead to significantly better pulses. However, we make a key observation which we term \textit{parameter monotonicity}. For both the VQE UCCSD and QAOA circuits, the appearances of $\theta_i$-dependent gates is monotonic in $i$--once a $\theta_i$ dependent gate appears, the subsequent parametrization-dependent gates must be $\theta_j$ for $j \geq i$. As a result, subcircuits with the same value of $\theta_i$ must be \emph{consecutive}. For example, the sequence of angles in parametrization-dependent gates could be [$\theta_1, \theta_1, \theta_2, \theta_3$] as in Figure~\ref{subfig:typical}, but not [$\theta_1, \theta_2, \theta_3, \theta_1$].

At a high level, parameter monotonicity for VQE/UCCSD and QAOA arise because their circuit constructions sequentially apply a circuit corresponding to each parameter exactly once. For instance, in QAOA, each parameter corresponds to the magnitude of Mixing or Cost-Optimization during the $i$th round--once the corresponding Mixing or Cost-Optimization has been applied, the circuit no longer depends on that parameter. Parameter monotonicity is not immediately obvious from visual inspection of variational circuits, because the circuit constructions and optimizations transform individual $\theta_i$-dependent gates to ones that are parametrized in terms $-\theta_i$ or $\theta_i/2$. We resolve these latent dependencies by explicitly tagging the dependent parameter in software during the variational circuit construction phase.

The implication of parameter monotonicity is that the subcircuits considered by flexible partial compilation are significantly deeper than the ones considered by strict partial compilation. Figure~\ref{subfig:flexible} demonstrates a small example; note that the $\theta_1$-dependent subcircuit indicated by red dashed lines is significantly deeper than the subcircuits generated by strict partial compilation.

\subsection{Hyperparameter Optimization}
\label{subsec:hyperparameter}

\begin{figure}
  \centering
    \includegraphics[width=0.5\textwidth]{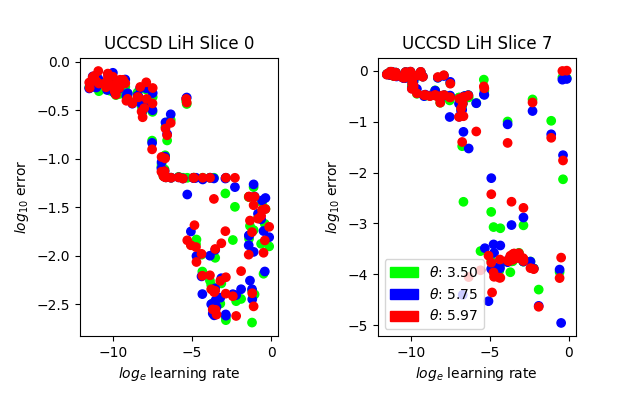}
      \caption{The 0\textsuperscript{th} single-angle dependent subcircuit of the UCCSD LiH circuit has two angle dependent gates, the 7\textsuperscript{th} has eight. These four qubit circuits are representative of the circuits studied in this work as well as larger future circuits due to the necessity of circuit blocking for circuits with more than four qubits. The graphs above plot GRAPE error against ADAM learning rate. For each permutation of the argument of the angle dependent gates in the subcircuits, the same range of learning rate values achieves the lowest error.}
      \label{fig:hyperparameter_plot}
\end{figure}

In GRAPE, an optimal control pulse is one that minimizes a set of cost functions corresponding to control amplitude, target state infidelity, and evolution time, among others\cite{NelsonPaper}. To obtain an optimal control pulse, the GRAPE algorithm manipulates a set of time-discrete control fields that act on a quantum system. It may analytically compute gradients of the cost functions to be minimized with respect to the control fields. These gradients are used to update control fields with an optimizer such as ADAM or L-BFGS-B. As opposed to the control fields, which are parameters manipulated by GRAPE, these optimizers have their own parameters such as learning rate and learning rate decay. These parameters are termed hyperparameters because they are set before the learning process begins.

Because they are inputs to the learning process, there is no closed form expression relating hyperparameters and the cost functions a learning model is minimizing. This makes hyperparameter optimization an ideal candidate for derivative free optimization techniques. Recent work has shown that tuning hyperparameters with methods such as bayesian optimization and radial basis functions can significantly improve performance for stochastic and expensive objectives such as minimizing the training error of neural networks \cite{thornton2013auto, diaz2017effective}. In our work, we employ hyperparameter optimization on GRAPE's ADAM optimizer. We realize faster convergence to a desired error rate over the baseline, significantly reducing compilation latency.

In particular, we make the observation that a high-performing hyperparameter configuration for a single-angle parameterized subcircuit is robust to changes in the argument of its $\theta_i$-dependent gates, as shown in Figure~\ref{fig:hyperparameter_plot}. Therefore, we are able to precompute high-performing hyperparameter configurations for each single-angle parameterized subcircuit and employ them in compilation. For each iteration of a variational algorithm, the argument of the $\theta_i$-dependent gates of each subcircuit will change, but the same hyperparameters are specified to GRAPE's optimizer, maintaining the same reduced compilation latency.

%% file: txt/8results.tex
\section{Results}
\label{sec:results}
\begin{figure}
  \centering
    \includegraphics[width=0.45\textwidth]{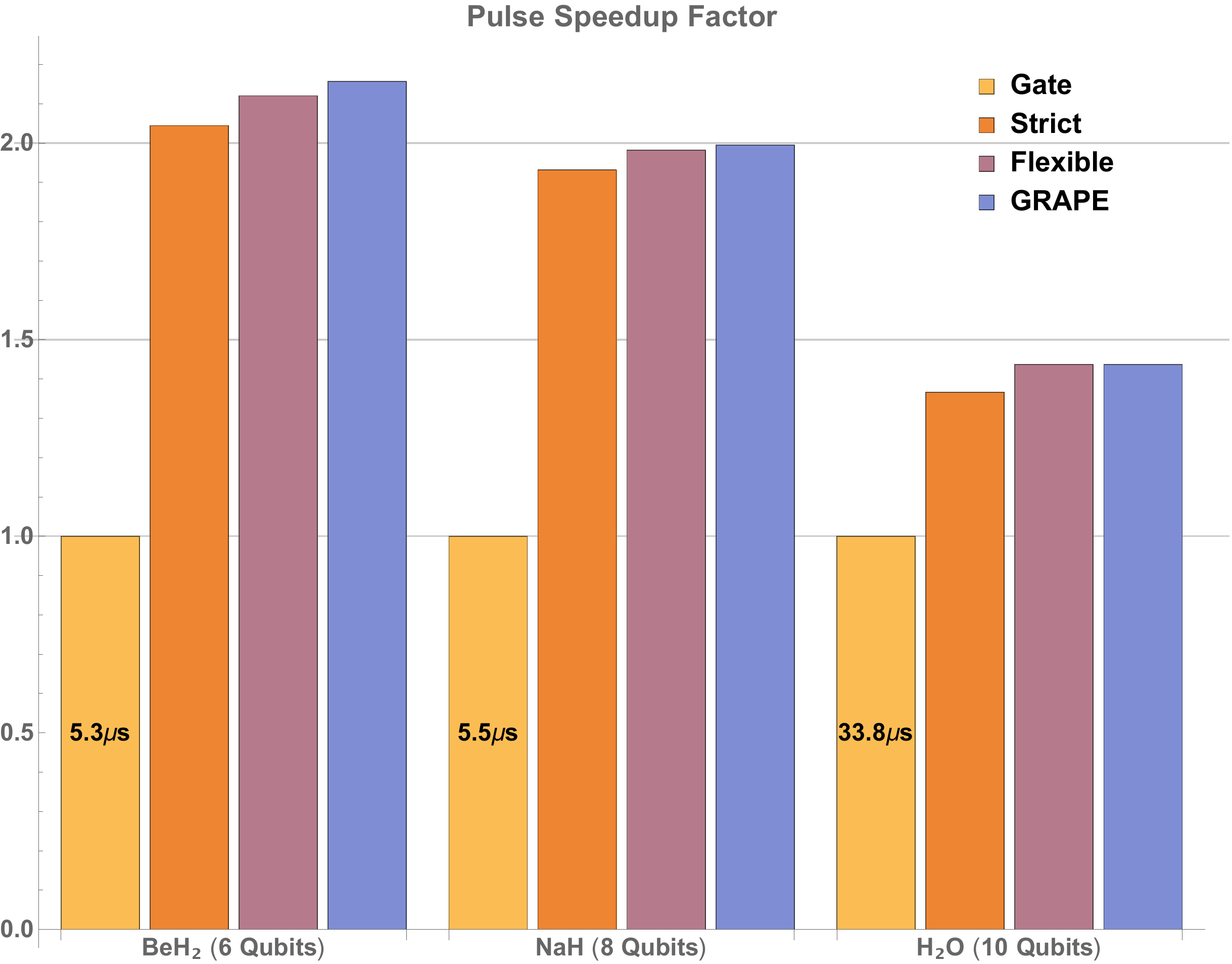}
      \caption{Pulse speedup factors (relative to gate based compilation) for VQE circuits. Full QOC 1.5-2x reductions in pulse durations for these circuits. Strict and flexible partial compilation recover 95\% and 99\% of this speedup respectively. Detailed results are reported in Table~\ref{tab:results}.}
      \label{fig:vqe_speedups}
\end{figure}

\begin{figure*}[ht]
\centering
    \begin{subfigure}{0.5\textwidth}
        \includegraphics[width=\textwidth]{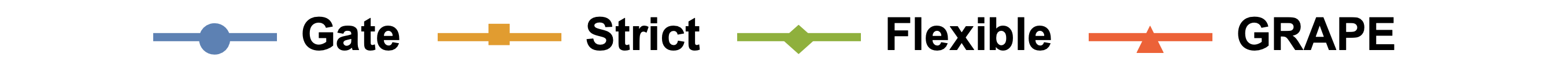}
    \end{subfigure}

\begin{subfigure}{0.44\textwidth}
        \includegraphics[width=\textwidth, trim=-.5cm -.5cm -1.5cm -.5cm]{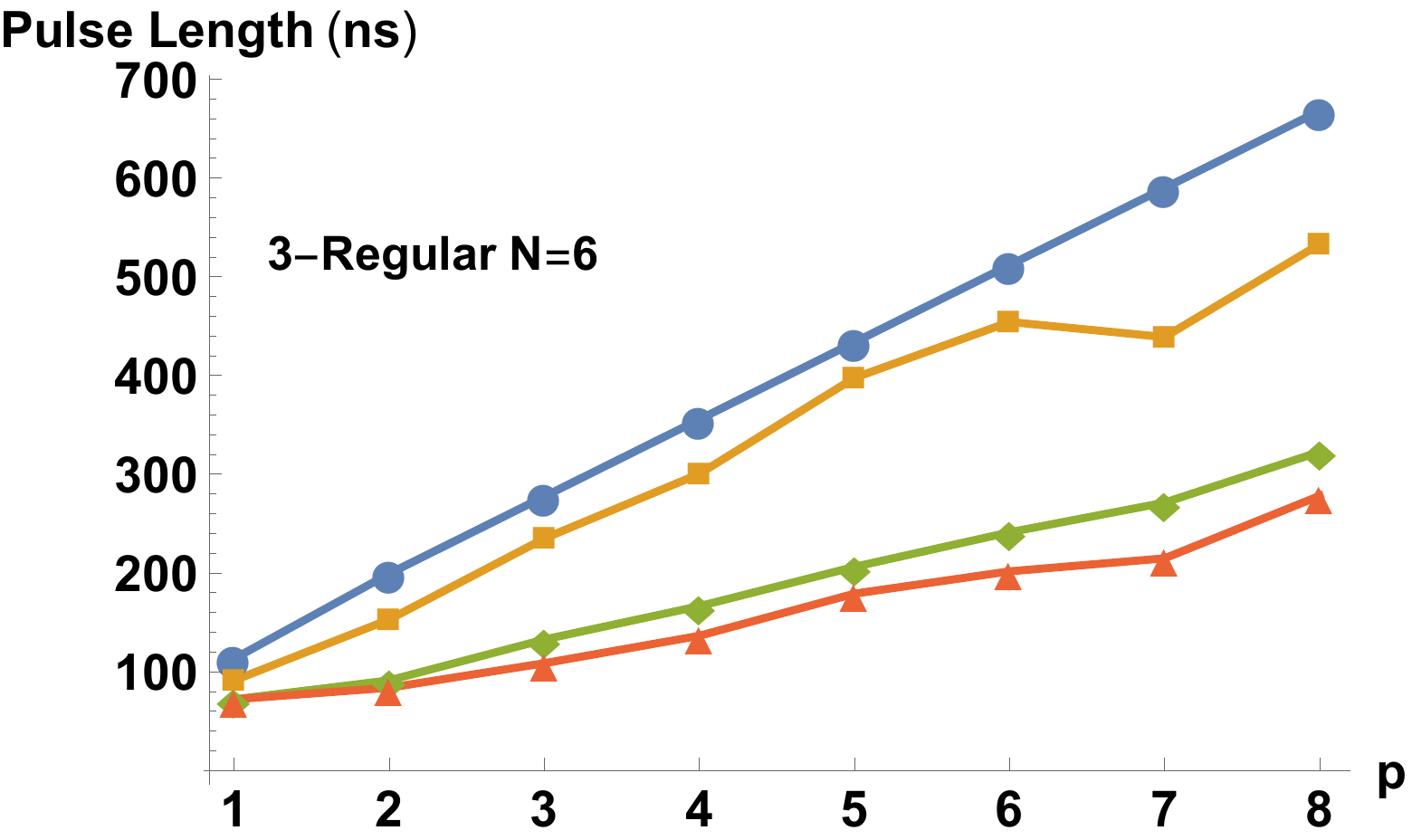}
    \end{subfigure}
    \begin{subfigure}{0.44\textwidth}
        \includegraphics[width=\textwidth, trim=-1.5cm -.5cm -.5cm -.5cm]{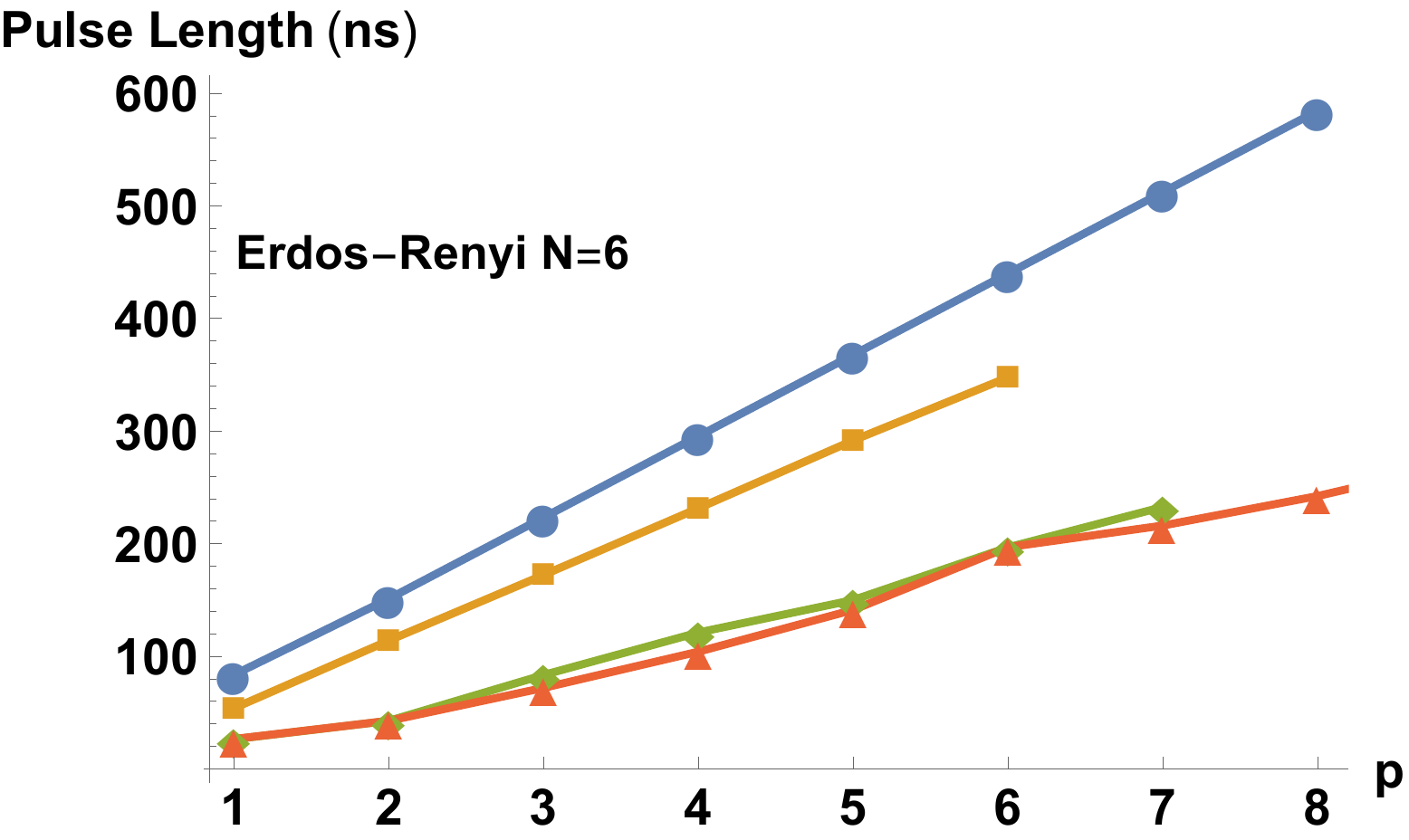}
    \end{subfigure}

\par\medskip

    \begin{subfigure}{0.44\textwidth}
        \includegraphics[width=\textwidth, trim=-.5cm -.5cm -1.5cm -.5cm]{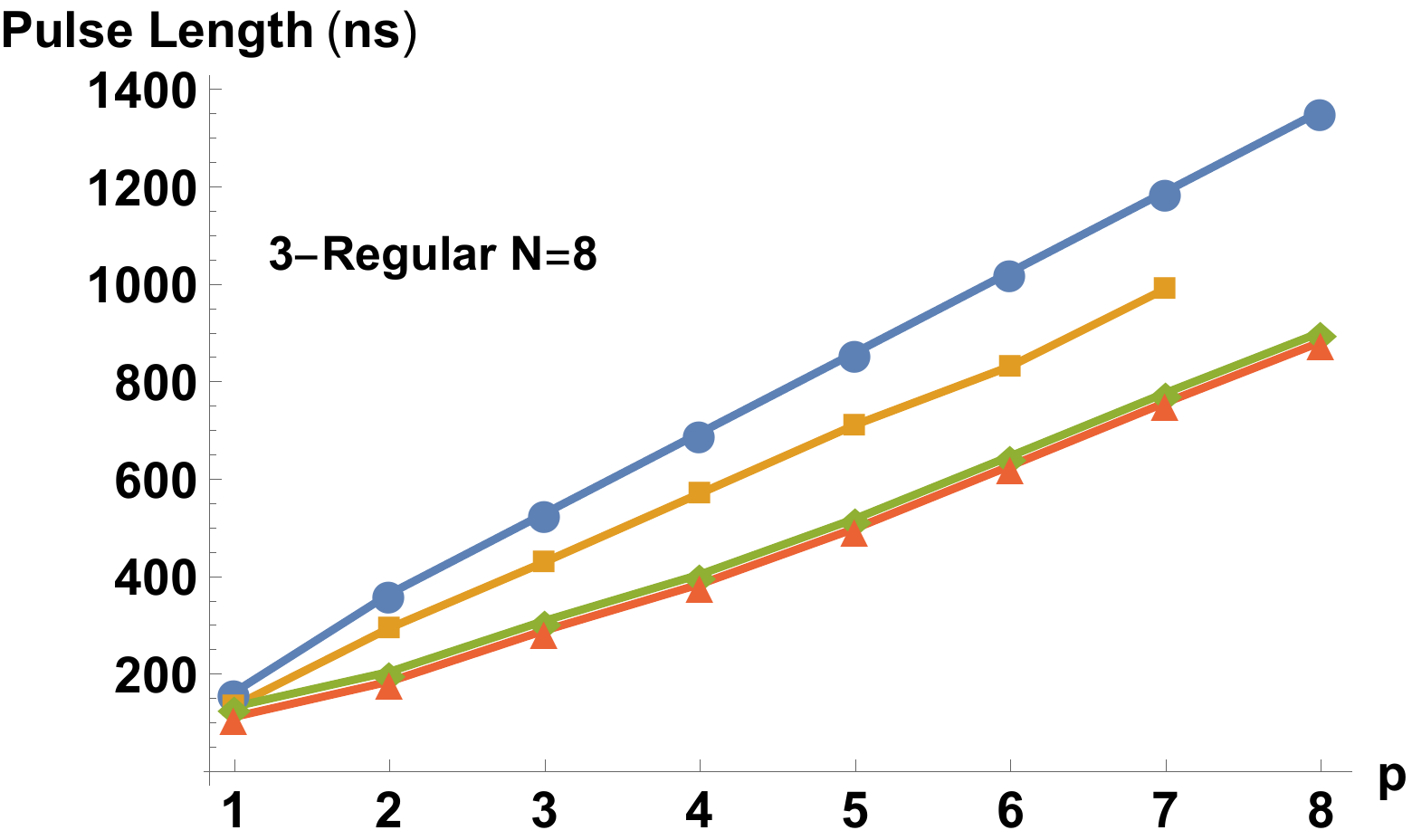}
    \end{subfigure}
    \begin{subfigure}{0.44\textwidth}
        \includegraphics[width=\textwidth, trim=-1.5cm -.5cm -.5cm -.5cm]{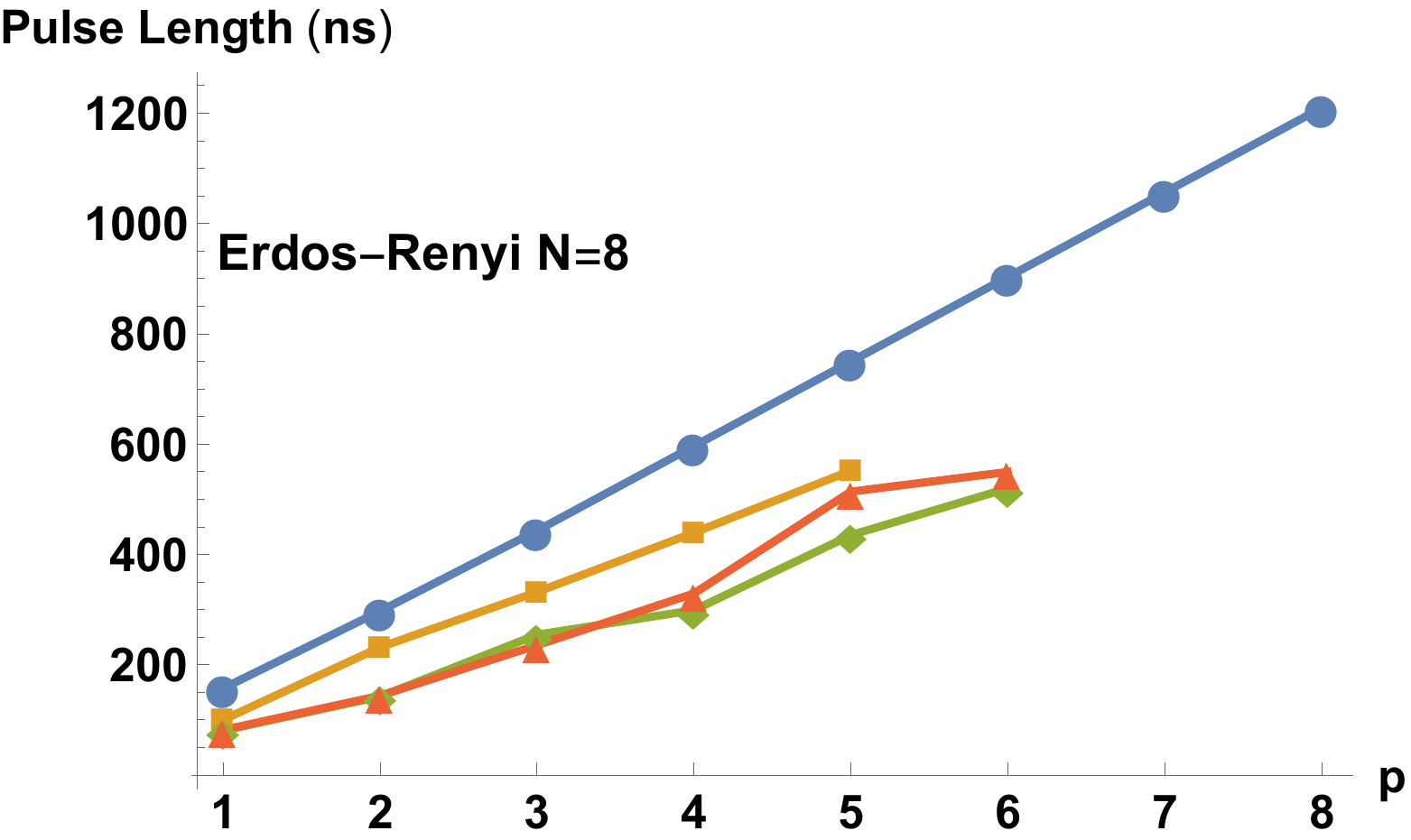}
    \end{subfigure}
    
    \caption{Pulse durations for QAOA MAXCUT benchmarks under the four compilation techniques, across all benchmarks. The gate based pulse time always increases linearly in $p$, the number of repeated rounds in the QAOA circuit. The average GRAPE pulse speedup is 2.6x for 6-node graphs and 1.8x for 8-node graphs. Strict partial compilation only achieves a modest speedup over gate based compilation, but flexible partial compilation essentially matches the GRAPE speedup exactly. The omitted data points correspond to computations that did not complete in 12 CPU-core hours, even after parallelizing subcircuit jobs.}
\label{fig:qaoa_speedups}
\end{figure*}

Our results were collected using over 200,000 CPU-core hours on Intel Xeon E5-2680 processors, using up to 64 GB of memory per GRAPE process. The large volume of compute is a result of both the high cost of running GRAPE and the number and large circuit size of benchmarks. We fixed randomization seeds when appropriate for both reproducability and consistency between identical benchmarks. Our results are available in Jupyter notebooks on our Github repository \cite{GithubRepo}.

\subsection{Pulse Speedups}

\begin{table*}[h!]
\scriptsize
\centering
\begin{tabular}{lccccccccccccccccc}
\hline\hline
 & & && & &&   \multicolumn{11}{c}{Max-Cut}\\
Compilation &\multicolumn{5}{c}{UCCSD} &&   \multicolumn{2}{c}{3-Regular, N=6} &&   \multicolumn{2}{c}{Erdos-Renyi, N=6}& &\multicolumn{2}{c}{3-Regular, N=8} && \multicolumn{2}{c}{Erdos-Renyi, N=8}\\
\cline{2-6}\cline{8-9}\cline{11-12}\cline{14-15}\cline{17-18}
Techniques   & H\textsubscript{2} &  LiH  &  BeH\textsubscript{2} & NaH &  H\textsubscript{2}O & & p=1  & p=5 && p=1 &  p=5 && p=1 & p=5 && p=1 & p=5\\
\hline
Gate-based & 35.3 & 871.1 & 5308.3 & 5490.4 & 33842.2 & & 113.2 & 433.6& & 83.7 & 367.8&&162.5& 860.0&&157.1 &749.5\\

Strict Partial & 15.0 & 307.0 & 2596.5 & 2842.7 & 24781.4 && 91.2&397.6&&54.0&291.8&&134.0&711.6&&100.0&551.7 \\

Flexible Partial & 5.0 & 84.0 & 2503.8 & 2770.8 & 23546.7 &&  72.0&206.2&&26.4&150.0&&112.0&498.9&&80.5&434.8 \\

Full GRAPE & 3.1 & 19.3 & 2461.7 & 2752.0 & 23546.7 &&  72.0&179.0&&26.6&141.2&&112.0&498.9&&81.6&513.7 \\

\hline\hline
\end{tabular}
\caption{Experimental results for pulse durations (in nanoseconds) across the VQE-UCCSD and QAOA benchmarks.}
\label{tab:results}
\end{table*}

Figure~\ref{fig:vqe_speedups} shows the pulse times speedup factors across our QAOA benchmarks for partial compilation and for full GRAPE, normalized to the gate-based compilation baseline. We present the normalized speedup factors, because the H\textsubscript{2}O VQE-UCCSD benchmark is 10x larger; the raw pulse times are presented in Table~\ref{tab:results}.

For the BeH\textsubscript{2} and NaH VQE-UCCSD benchmarks, full GRAPE gives a 2.15x and 2.00x speedup in pulse duration respectively. Strict partial compilation is able to recover almost this full advantage, with speedups at 2.04x and 1.93x respectively. As discussed in Section~\ref{sec:strict_partial}, this matches the expectations, because the VQE-UCCSD benchmarks have relatively deep Fixed subcircuits. Finally, the speedups for flexible partial compilation are 2.12x and 1.98x, which nearly closes the gap between strict partial compilation and GRAPE.

The H\textsubscript{2}O benchmark has similar relative speedups between strict, flexible, and GRAPE, with factors of 1.37, 1.44, 1.44.\footnote{In fact, the pulse speedup for flexible partial compilation exactly matches GRAPE, because each 4-qubit block handled by GRAPE depends on at most one parameter.} However, the advantage over gate based compilation is smaller than for the BeH\textsubscript{2} and NaH benchmarks.

\begin{figure}[ht]
  \centering
    \includegraphics[width=0.4\textwidth]{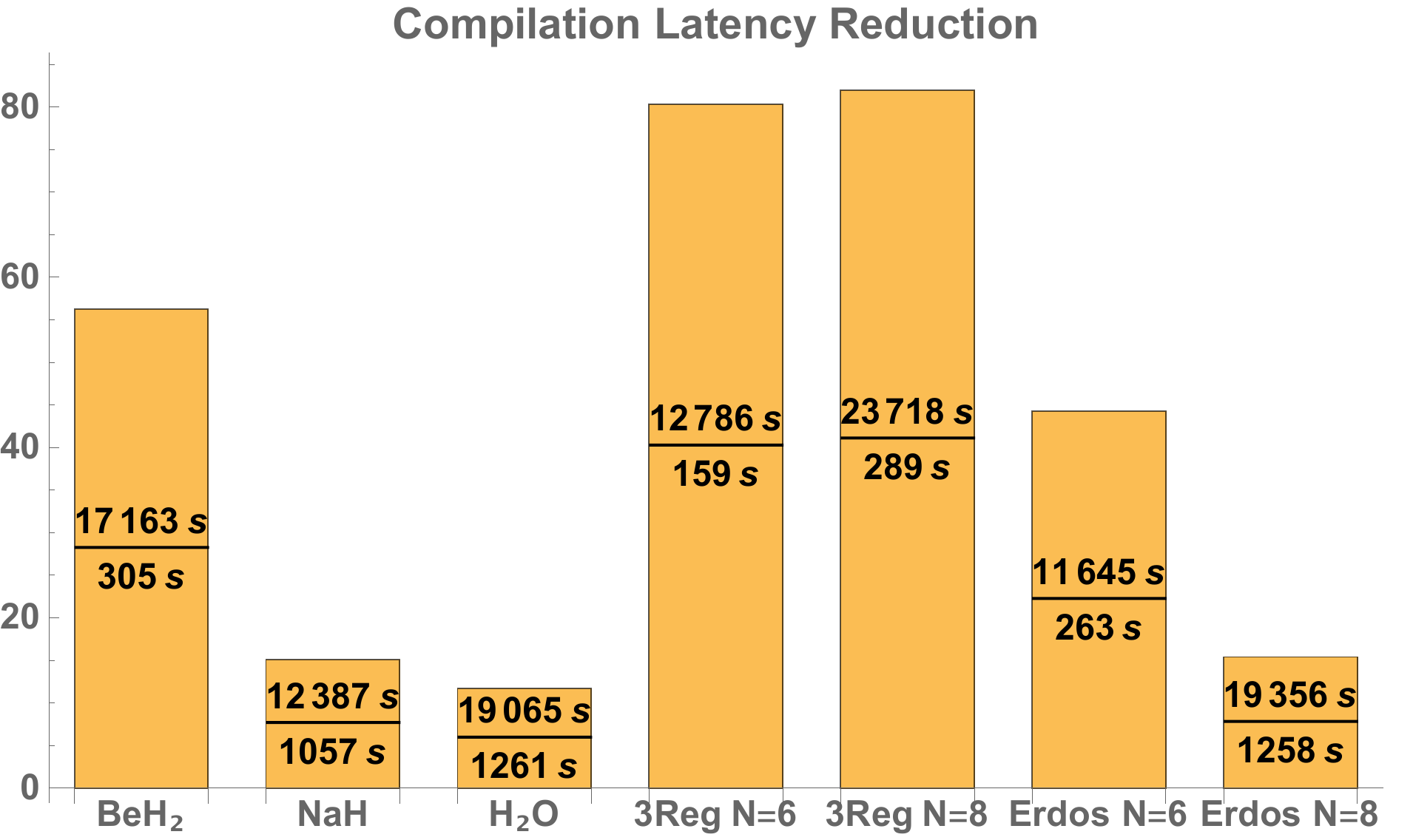}
      \caption{Reduction factors in compilation latency. The ratios indicate the average compilation latency using flexible partial compilation divided by latency using full GRAPE compilation. Flexible partial compilation uses about an hour of pre-compute time to determine the best learning rate - decay rate pair for each subcircuit.}
      \label{fig:compilation_latency_speedups}
\end{figure}

Figure~\ref{fig:qaoa_speedups} shows results for QAOA benchmarks. Strict partial compilation has speedups of 1.22x and 1.33x across the $N=6$ and $N=8$ qubit benchmarks respectively. By contrast, flexible partial compilation has average speedups of 2.3x and 1.8x across the $N=6$ and $N=8$ benchmarks, which almost matches the results from GRAPE. This separation between strict and flexible partial compilation matches the expected results discussed in Section~\ref{sec:strict_partial}. The high frequency of parametrized gates in QAOA limits the depth of Fixed blocks, so strict blocking has limited mileage. However, due to the four-qubit maximum subcircuit size for GRAPE, each block will rarely depend on more than one $\theta_i$ parameter. On these single-angle dependent blocks, flexible partial compilation achieves the same pulse speedups as GRAPE.

\subsection{Compilation Latency Reduction}

Figure~\ref{fig:compilation_latency_speedups} shows the compilation latency reduction achieved by flexible partial compilation, relative to full GRAPE compilation. As described in Section 7.2, flexible partial compilation is able to dramatically speed up the gradient descent's convergence by tuning the learning rate and decay rate hyperparameters on a per-subcircuit basis. We note that the 3-regular graphs achieve particularly high compilation latency reduction factors of 80.3x and 81.9x. Across all benchmarks, the reduction in compile time is from hours to minutes, which is critical in the context of variational algorithms.

\subsection{Simulation with Realistic Pulses}
While we performed our GRAPE runs without accounting for error or noise sources for simplicity, it can be adapted to account for these sources. For example, we could demand well-shaped pulses, account for leakage into higher states outside the binary qubit abstraction, or explicitly model the qubit decoherence times. To demonstrate that these  sources can be accounted for, we re-ran two of our VQE and QAOA benchmarks with Full GRAPE using these more realistic assumptions:
\begin{itemize}
    \item Allowing only 1 pulse datapoint every nanosecond (1 GSa/s), versus 20 GSa/s in the other results presented in this paper.
    \item Including leakage into the qutrit \textit{leakage} level. Other results in this paper use the binary-qubit approximation, as outlined in the system Hamiltonian in Appendix~\ref{sec:experimental_setup}.
    \item Application of aggressive pulse regularization in GRAPE to ensure that the pulse shapes follow a Gaussian envelope and have smooth 1st and 2nd derivatives.
\end{itemize}

Table~\ref{tab:results} compares the pulse speedups due to GRAPE, under both our standard (less realistic) GRAPE settings and under the more realistic settings that account for the three items above. For VQE and QAOA applications, the GRAPE speedups speedups are 11.4x (standard) vs 8.8x (realistic) and 4.5x (standard) vs. 3.0x (realistic) respectively. While the more realistic pulses do seem to have somewhat lower pulse speedup factors, they are similar and still feature significant speedup over gate based compilation.

\renewcommand{\arraystretch}{1.4} 
\begin{table}[h]
\small
\centering
\begin{tabular}{|c|c|c|ll}
\cline{1-3}
Gate ns $\rightarrow$ GRAPE ns (reduction) & H\textsubscript{2} VQE & Erdos-Renyi $N=3$ &  \\ \cline{1-3}
Standard    & 35.3 $\rightarrow$ 3.1 (11.4x)      & 15.0 $\rightarrow$ 3.3 (4.5x)     &  \\ \cline{1-3}
More Realistic                   & 420 $\rightarrow$ 48 (8.8x)      & 285 $\rightarrow$ 96 (3.0x)    &  \\ \cline{1-3}
\end{tabular}
\caption{Speedups due to GRAPE compilation under the standard settings and under more realistic settings, which account for lower sampling rates, qutrit leakage, and pulse regularization. Results are given for the H\textsubscript{2} VQE benchmark and for the Erdos-Renyi $N=3$ QAOA benchmark. The speedup factors due to GRAPE are similar with and without the more realistic assumptions.}
\label{tab:realstic}
\end{table}

\subsection{Aggregate Impact on Total Runtime}
For a quantitative sense of aggregate impact, we note that VQE requires thousands of iterations, even for small molecules. For instance, past work in VQE, towards estimating the ground state energy of BeH\textsubscript{2}, required 3500 iterations \cite{Kandala}. Per Figure~\ref{fig:compilation_latency_speedups}, this would amount to over 2 years of runtime compilation latency via Full-GRAPE. By contrast, strict partial compilation achieves zero runtime compilation latency via lookup table, and the pre-computed pulses for the fixed blocks were compiled in under 1 hour. Since the UCCSD ansatz has quartic scaling in the number of parameters \cite{UCCSD_Strategies}, the number of iterations required scales aggressively for bigger molecules and the advantage of our approach will scale favorably. Further experimental work is needed to estimate the advantage of our approach for larger molecules in terms of total runtime, but extrapolation from small molecules BeH\textsubscript{2} seems promising. Similarly, while the improvement in quality-of-result due to the shorter pulse times from GRAPE is difficult to quantify without concrete experiments, we emphasize that the error due to decoherence scales exponentially with quantum runtime. Therefore, we again expect favorable results, owing to the significant pulse time speedup of our techniques relative to gate based compilation.

%% file: txt/9conclusion.tex
\section{Conclusion}
\label{sec:conclusion}
Variational quantum algorithms such as VQE and QAOA are strong candidates for demonstrating a quantum advantage in problems such as molecular ground state estimation, MAXCUT approximation, and prime factorization. Unlike prior algorithms, variational algorithms are parametrized, with the parameters at each iteration determined based on the results of previous iterations. Consequently, compilation is interleaved with computation. As a result, it is not practical to each variational circuit with out-of-the-box GRAPE, which takes several minutes to find an optimized pulse even on small (4-qubit) circuit.

Our partial compilation techniques offer a path to achieving the pulse speedups of GRAPE, without incurring its compilation latency. On the VQE-UCCSD circuits, our strict partial compilation strategy achieves 1.5x-2x pulse speedups over gate based compilation, almost matching the speedups from full GRAPE. Strict partial compilation is performed by precomputing optimal pulses for Fixed blocks. During execution, it has the same--essentially instant--lookup table based compilation procedure as gate based compilation. Thus, strict partial compilation is strictly better than gate based compilation.

For QAOA circuits, while strict partial compilation only achieves modest pulse speedup, we find that flexible partial compilation almost exactly matches the pulse speedups of GRAPE. Flexible partial compilation precomputes the best hyperparameters for each slice, so that when the $\theta_i$ parameters are specified at runtime, an optimized pulse sequence can be computed rapidly. For our benchmarked circuits, we found 10-100x reductions in compilation latency from flexible partial compilation, relative to full GRAPE compilation.

We emphasize that achieving optimized pulses is critical because error due to decoherence error is exponential in the pulse duration. Thus, our pulse speedups are not merely about wall time speedups for quantum circuits, but moreso about making computations possible in the first place, before the qubits decohere.

\section{Future Work}
\label{sec:future_work}
The industry adoption of the OpenPulse standard will usher an experimental era for pulse-level control. Running our partial compilation schemes on an actual machine will be valuable in terms of determining exactly how to weigh tradeoffs between pre-computation resources, compilation latency, and pulse durations.

On the computational side, we also see significant potential for extending the scalability of GRAPE. While past work has successfully used GRAPE on 10 qubit widths with very simple circuits (for example, 10 identical single-qubit rotations in parallel), we found that for complicated circuits, GRAPE only converges reliabily with widths up to 4 qubits. This 4-qubit blocking width limits the depths of the subcircuits that both GRAPE and our partial compilation schemes can consider. For example, in the additional two VQE-UCCSD molecules benchmarks (H\textsubscript{2} and LiH) reported in Table 4, flexible partial compilation and full GRAPE achieve 7-50x pulse speedups because the benchmarks are 2 and 4 qubits in width. Thus, investigating the convergence properties of GRAPE and extending the circuit widths it reliably converges for will substantially extend the advantage that these techniques can achieve over gate based compilation. 

%% file: txt/4experimental_setup.tex
\section{System Hamiltonian} \label{sec:experimental_setup}
Although our techniques are general and apply to any quantum computer, the pulses produced by GRAPE are specific to the underlying hardware platform. We chose to compile to control pulses for a quantum computer with gmon superconducting qubits \cite{Gmon}, because this qubit type is one of the leader contenders for scalable quantum machines. For instance, the gmon qubit is central to Google's experimental efforts for demonstrating quantum supremacy.

The control pulse inputs that we specified to GRAPE were based on the gmon's system Hamiltonian. Each qubit, $j$, has a flux-drive control pulse and a charge-drive control pulse which have respective Hamiltonians, truncated to the qubit subspace:
$$H_{c,j}(t) =  \sum_{j=1}^N \Omega_{c, j}(t) (a_j^\dagger + a_j) = \sum_{j=1}^N \Omega_{c, j}(t)  \begin{pmatrix} 0 & 1 \\ 1 & 0 \end{pmatrix} $$
and
$$H_{f,j}(t) = \sum_{j=1}^N \Omega_{f, j}(t)(a_j^\dagger a_j) = \sum_{j=1}^N \Omega_{f, j} \begin{pmatrix} 0 & 0 \\ 0 & 1 \end{pmatrix}$$
It can be seen from exponentiating these matrices that the control pulses correspond to $R_x(\theta)$ and $R_z(\phi)$ type gates respectively. We chose maximium drive amplitudes of $|\Omega_{c, j}(t)| \leq 2 \pi \times 0.1$ GHz and $|\Omega_{f, j}(t)| \leq 2 \pi \times 1.5$ GHz. These values, including the asymmetry between charge and flux drive, are representative of typical machines.

In addition to these single qubit terms, there is a control pulse for each pair of connected qubits. We consider a rectangular-grid topology with nearest-neighbor connectivity. Between each connected pair of qubits $j$ and $k$, the corresponding control Hamiltonian is
$$ H_{j,k}(t) = g(t) (a_j^\dagger + a_j)(a_k^\dagger + a_k)$$
This two-qubit interaction type corresponds to the entangling iSWAP gate (which swaps two qubits and also applies a phase factor). We use a maximum coupling strength of $|g(t)| \leq 2 \pi \times 50$ MHz

Within the GRAPE software, we discretized the control pulses to 0.05 ns time slices. We set a target fidelity of 99.9\% for each invocation of GRAPE. Raw data from all of our GRAPE runs are available at our Github repository \cite{GithubRepo}.

%% file: main.bbl

\begin{thebibliography}{53}


\ifx \showCODEN    \undefined \def \showCODEN     #1{\unskip}     \fi
\ifx \showDOI      \undefined \def \showDOI       #1{#1}\fi
\ifx \showISBNx    \undefined \def \showISBNx     #1{\unskip}     \fi
\ifx \showISBNxiii \undefined \def \showISBNxiii  #1{\unskip}     \fi
\ifx \showISSN     \undefined \def \showISSN      #1{\unskip}     \fi
\ifx \showLCCN     \undefined \def \showLCCN      #1{\unskip}     \fi
\ifx \shownote     \undefined \def \shownote      #1{#1}          \fi
\ifx \showarticletitle \undefined \def \showarticletitle #1{#1}   \fi
\ifx \showURL      \undefined \def \showURL       {\relax}        \fi
\providecommand\bibfield[2]{#2}
\providecommand\bibinfo[2]{#2}
\providecommand\natexlab[1]{#1}
\providecommand\showeprint[2][]{arXiv:#2}

\bibitem[\protect\citeauthoryear{Abdelhafez, Schuster, and Koch}{Abdelhafez
  et~al\mbox{.}}{2019}]%
        {MohamedPaper}
\bibfield{author}{\bibinfo{person}{Mohamed Abdelhafez},
  \bibinfo{person}{David~I. Schuster}, {and} \bibinfo{person}{Jens Koch}.}
  \bibinfo{year}{2019}\natexlab{}.
\newblock \bibinfo{title}{Gradient-based optimal control of open quantum
  systems using quantum trajectories and automatic differentiation}.
\newblock
\newblock
\showeprint{arXiv:1901.05541}


\bibitem[\protect\citeauthoryear{Anschuetz, Olson, Aspuru-Guzik, and
  Cao}{Anschuetz et~al\mbox{.}}{2018}]%
        {VQF}
\bibfield{author}{\bibinfo{person}{Eric~R. Anschuetz},
  \bibinfo{person}{Jonathan~P. Olson}, \bibinfo{person}{Al{\'a}n Aspuru-Guzik},
  {and} \bibinfo{person}{Yudong Cao}.} \bibinfo{year}{2018}\natexlab{}.
\newblock \bibinfo{title}{Variational Quantum Factoring}.
\newblock
\newblock
\showeprint{arXiv:1808.08927}


\bibitem[\protect\citeauthoryear{Babbush, McClean, Wecker, Aspuru-Guzik, and
  Wiebe}{Babbush et~al\mbox{.}}{2015}]%
        {Babbush2015}
\bibfield{author}{\bibinfo{person}{Ryan Babbush}, \bibinfo{person}{Jarrod
  McClean}, \bibinfo{person}{Dave Wecker}, \bibinfo{person}{Al\'an
  Aspuru-Guzik}, {and} \bibinfo{person}{Nathan Wiebe}.}
  \bibinfo{year}{2015}\natexlab{}.
\newblock \showarticletitle{Chemical basis of Trotter-Suzuki errors in quantum
  chemistry simulation}.
\newblock \bibinfo{journal}{\emph{Phys. Rev. A}}  \bibinfo{volume}{91}
  (\bibinfo{date}{Feb} \bibinfo{year}{2015}), \bibinfo{pages}{022311}.
\newblock
Issue 2.
\urldef\tempurl%
\url{https://doi.org/10.1103/PhysRevA.91.022311}
\showDOI{\tempurl}


\bibitem[\protect\citeauthoryear{Barends, Shabani, Lamata, Kelly, Mezzacapo,
  Heras, Babbush, Fowler, Campbell, Chen, Chen, Chiaro, Dunsworth, Jeffrey,
  Lucero, Megrant, Mutus, Neeley, Neill, O'Malley, Quintana, Roushan, Sank,
  Vainsencher, Wenner, White, Solano, Neven, and Martinis}{Barends
  et~al\mbox{.}}{2016}]%
        {DigitizedAdiabaticQC}
\bibfield{author}{\bibinfo{person}{R. Barends}, \bibinfo{person}{A. Shabani},
  \bibinfo{person}{L. Lamata}, \bibinfo{person}{J. Kelly}, \bibinfo{person}{A.
  Mezzacapo}, \bibinfo{person}{U.~Las Heras}, \bibinfo{person}{R. Babbush},
  \bibinfo{person}{A.~G. Fowler}, \bibinfo{person}{B. Campbell},
  \bibinfo{person}{Yu Chen}, \bibinfo{person}{Z. Chen}, \bibinfo{person}{B.
  Chiaro}, \bibinfo{person}{A. Dunsworth}, \bibinfo{person}{E. Jeffrey},
  \bibinfo{person}{E. Lucero}, \bibinfo{person}{A. Megrant},
  \bibinfo{person}{J.~Y. Mutus}, \bibinfo{person}{M. Neeley},
  \bibinfo{person}{C. Neill}, \bibinfo{person}{P.~J.~J. O'Malley},
  \bibinfo{person}{C. Quintana}, \bibinfo{person}{P. Roushan},
  \bibinfo{person}{D. Sank}, \bibinfo{person}{A. Vainsencher},
  \bibinfo{person}{J. Wenner}, \bibinfo{person}{T.~C. White},
  \bibinfo{person}{E. Solano}, \bibinfo{person}{H. Neven}, {and}
  \bibinfo{person}{John~M. Martinis}.} \bibinfo{year}{2016}\natexlab{}.
\newblock \showarticletitle{Digitized adiabatic quantum computing with a
  superconducting circuit}.
\newblock \bibinfo{journal}{\emph{Nature}}  \bibinfo{volume}{534}
  (\bibinfo{date}{08 Jun} \bibinfo{year}{2016}), \bibinfo{pages}{222 EP --}.
\newblock
\urldef\tempurl%
\url{https://doi.org/10.1038/nature17658}
\showURL{%
\tempurl}


\bibitem[\protect\citeauthoryear{Barkoutsos, Gonthier, Sokolov, Moll, Salis,
  Fuhrer, Ganzhorn, Egger, Troyer, Mezzacapo, Filipp, and
  Tavernelli}{Barkoutsos et~al\mbox{.}}{2018}]%
        {IBM_UCCSD}
\bibfield{author}{\bibinfo{person}{Panagiotis~Kl. Barkoutsos},
  \bibinfo{person}{Jerome~F. Gonthier}, \bibinfo{person}{Igor Sokolov},
  \bibinfo{person}{Nikolaj Moll}, \bibinfo{person}{Gian Salis},
  \bibinfo{person}{Andreas Fuhrer}, \bibinfo{person}{Marc Ganzhorn},
  \bibinfo{person}{Daniel~J. Egger}, \bibinfo{person}{Matthias Troyer},
  \bibinfo{person}{Antonio Mezzacapo}, \bibinfo{person}{Stefan Filipp}, {and}
  \bibinfo{person}{Ivano Tavernelli}.} \bibinfo{year}{2018}\natexlab{}.
\newblock \showarticletitle{Quantum algorithms for electronic structure
  calculations: Particle-hole Hamiltonian and optimized wave-function
  expansions}.
\newblock \bibinfo{journal}{\emph{Phys. Rev. A}}  \bibinfo{volume}{98}
  (\bibinfo{date}{Aug} \bibinfo{year}{2018}), \bibinfo{pages}{022322}.
\newblock
Issue 2.
\urldef\tempurl%
\url{https://doi.org/10.1103/PhysRevA.98.022322}
\showDOI{\tempurl}


\bibitem[\protect\citeauthoryear{Bartlett and Musia\l{}}{Bartlett and
  Musia\l{}}{2007}]%
        {RevModPhys.79.291}
\bibfield{author}{\bibinfo{person}{Rodney~J. Bartlett} {and}
  \bibinfo{person}{Monika Musia\l{}}.} \bibinfo{year}{2007}\natexlab{}.
\newblock \showarticletitle{Coupled-cluster theory in quantum chemistry}.
\newblock \bibinfo{journal}{\emph{Rev. Mod. Phys.}}  \bibinfo{volume}{79}
  (\bibinfo{date}{Feb} \bibinfo{year}{2007}), \bibinfo{pages}{291--352}.
\newblock
Issue 1.
\urldef\tempurl%
\url{https://doi.org/10.1103/RevModPhys.79.291}
\showDOI{\tempurl}


\bibitem[\protect\citeauthoryear{Chen, Neill, Roushan, Leung, Fang, Barends,
  Kelly, Campbell, Chen, Chiaro, Dunsworth, Jeffrey, Megrant, Mutus, O'Malley,
  Quintana, Sank, Vainsencher, Wenner, White, Geller, Cleland, and
  Martinis}{Chen et~al\mbox{.}}{2014}]%
        {Gmon}
\bibfield{author}{\bibinfo{person}{Yu Chen}, \bibinfo{person}{C. Neill},
  \bibinfo{person}{P. Roushan}, \bibinfo{person}{N. Leung}, \bibinfo{person}{M.
  Fang}, \bibinfo{person}{R. Barends}, \bibinfo{person}{J. Kelly},
  \bibinfo{person}{B. Campbell}, \bibinfo{person}{Z. Chen}, \bibinfo{person}{B.
  Chiaro}, \bibinfo{person}{A. Dunsworth}, \bibinfo{person}{E. Jeffrey},
  \bibinfo{person}{A. Megrant}, \bibinfo{person}{J.~Y. Mutus},
  \bibinfo{person}{P.~J.~J. O'Malley}, \bibinfo{person}{C.~M. Quintana},
  \bibinfo{person}{D. Sank}, \bibinfo{person}{A. Vainsencher},
  \bibinfo{person}{J. Wenner}, \bibinfo{person}{T.~C. White},
  \bibinfo{person}{Michael~R. Geller}, \bibinfo{person}{A.~N. Cleland}, {and}
  \bibinfo{person}{John~M. Martinis}.} \bibinfo{year}{2014}\natexlab{}.
\newblock \showarticletitle{Qubit Architecture with High Coherence and Fast
  Tunable Coupling}.
\newblock \bibinfo{journal}{\emph{Phys. Rev. Lett.}}  \bibinfo{volume}{113}
  (\bibinfo{date}{Nov} \bibinfo{year}{2014}), \bibinfo{pages}{220502}.
\newblock
Issue 22.
\urldef\tempurl%
\url{https://doi.org/10.1103/PhysRevLett.113.220502}
\showDOI{\tempurl}


\bibitem[\protect\citeauthoryear{Chou, Huang, and Goan}{Chou
  et~al\mbox{.}}{2015}]%
        {PhysRevA.91.052315}
\bibfield{author}{\bibinfo{person}{Yi Chou}, \bibinfo{person}{Shang-Yu Huang},
  {and} \bibinfo{person}{Hsi-Sheng Goan}.} \bibinfo{year}{2015}\natexlab{}.
\newblock \showarticletitle{Optimal control of fast and high-fidelity quantum
  gates with electron and nuclear spins of a nitrogen-vacancy center in
  diamond}.
\newblock \bibinfo{journal}{\emph{Phys. Rev. A}}  \bibinfo{volume}{91}
  (\bibinfo{date}{May} \bibinfo{year}{2015}), \bibinfo{pages}{052315}.
\newblock
Issue 5.
\urldef\tempurl%
\url{https://doi.org/10.1103/PhysRevA.91.052315}
\showDOI{\tempurl}


\bibitem[\protect\citeauthoryear{Crooks}{Crooks}{2018}]%
        {Rigetti_QAOA_Simulation}
\bibfield{author}{\bibinfo{person}{Gavin~E. Crooks}.}
  \bibinfo{year}{2018}\natexlab{}.
\newblock \bibinfo{title}{Performance of the Quantum Approximate Optimization
  Algorithm on the Maximum Cut Problem}.
\newblock
\newblock
\showeprint{arXiv:1811.08419}


\bibitem[\protect\citeauthoryear{{de Fouquieres}, {Schirmer}, {Glaser}, and
  {Kuprov}}{{de Fouquieres} et~al\mbox{.}}{2011}]%
        {GRAPE2}
\bibfield{author}{\bibinfo{person}{P. {de Fouquieres}}, \bibinfo{person}{S.~G.
  {Schirmer}}, \bibinfo{person}{S.~J. {Glaser}}, {and} \bibinfo{person}{I.
  {Kuprov}}.} \bibinfo{year}{2011}\natexlab{}.
\newblock \showarticletitle{{Second order gradient ascent pulse engineering}}.
\newblock \bibinfo{journal}{\emph{Journal of Magnetic Resonance}}
  \bibinfo{volume}{212} (\bibinfo{date}{Oct.} \bibinfo{year}{2011}),
  \bibinfo{pages}{412--417}.
\newblock
\urldef\tempurl%
\url{https://doi.org/10.1016/j.jmr.2011.07.023}
\showDOI{\tempurl}
\showeprint[arxiv]{quant-ph/1102.4096}


\bibitem[\protect\citeauthoryear{Diaz, Fokoue-Nkoutche, Nannicini, and
  Samulowitz}{Diaz et~al\mbox{.}}{2017}]%
        {diaz2017effective}
\bibfield{author}{\bibinfo{person}{Gonzalo~I Diaz}, \bibinfo{person}{Achille
  Fokoue-Nkoutche}, \bibinfo{person}{Giacomo Nannicini}, {and}
  \bibinfo{person}{Horst Samulowitz}.} \bibinfo{year}{2017}\natexlab{}.
\newblock \showarticletitle{An effective algorithm for hyperparameter
  optimization of neural networks}.
\newblock \bibinfo{journal}{\emph{IBM Journal of Research and Development}}
  \bibinfo{volume}{61}, \bibinfo{number}{4/5} (\bibinfo{year}{2017}),
  \bibinfo{pages}{9--1}.
\newblock


\bibitem[\protect\citeauthoryear{Dolde, Bergholm, Wang, Jakobi, Naydenov,
  Pezzagna, Meijer, Jelezko, Neumann, Schulte-Herbr{\"u}ggen, Biamonte, and
  Wrachtrup}{Dolde et~al\mbox{.}}{2014}]%
        {QOC_Crosstalk}
\bibfield{author}{\bibinfo{person}{Florian Dolde}, \bibinfo{person}{Ville
  Bergholm}, \bibinfo{person}{Ya Wang}, \bibinfo{person}{Ingmar Jakobi},
  \bibinfo{person}{Boris Naydenov}, \bibinfo{person}{S{\'e}bastien Pezzagna},
  \bibinfo{person}{Jan Meijer}, \bibinfo{person}{Fedor Jelezko},
  \bibinfo{person}{Philipp Neumann}, \bibinfo{person}{Thomas
  Schulte-Herbr{\"u}ggen}, \bibinfo{person}{Jacob Biamonte}, {and}
  \bibinfo{person}{J{\"o}rg Wrachtrup}.} \bibinfo{year}{2014}\natexlab{}.
\newblock \showarticletitle{High-fidelity spin entanglement using optimal
  control}.
\newblock \bibinfo{journal}{\emph{Nature Communications}}  \bibinfo{volume}{5}
  (\bibinfo{date}{28 Feb} \bibinfo{year}{2014}), \bibinfo{pages}{3371 EP --}.
\newblock
\urldef\tempurl%
\url{https://doi.org/10.1038/ncomms4371}
\showURL{%
\tempurl}
\newblock
\shownote{Article.}


\bibitem[\protect\citeauthoryear{Eyring}{Eyring}{1935}]%
        {Eyring}
\bibfield{author}{\bibinfo{person}{Henry Eyring}.}
  \bibinfo{year}{1935}\natexlab{}.
\newblock \showarticletitle{The Activated Complex in Chemical Reactions}.
\newblock \bibinfo{journal}{\emph{The Journal of Chemical Physics}}
  \bibinfo{volume}{3}, \bibinfo{number}{2} (\bibinfo{year}{1935}),
  \bibinfo{pages}{107--115}.
\newblock
\urldef\tempurl%
\url{https://doi.org/10.1063/1.1749604}
\showDOI{\tempurl}
\showeprint{https://doi.org/10.1063/1.1749604}


\bibitem[\protect\citeauthoryear{Farhi, Goldstone, and Gutmann}{Farhi
  et~al\mbox{.}}{2014}]%
        {QAOA}
\bibfield{author}{\bibinfo{person}{Edward Farhi}, \bibinfo{person}{Jeffrey
  Goldstone}, {and} \bibinfo{person}{Sam Gutmann}.}
  \bibinfo{year}{2014}\natexlab{}.
\newblock \bibinfo{title}{A Quantum Approximate Optimization Algorithm}.
\newblock
\newblock
\showeprint{arXiv:1411.4028}


\bibitem[\protect\citeauthoryear{Farhi and Harrow}{Farhi and Harrow}{2016}]%
        {QAOA_Supremacy}
\bibfield{author}{\bibinfo{person}{Edward Farhi} {and} \bibinfo{person}{Aram~W
  Harrow}.} \bibinfo{year}{2016}\natexlab{}.
\newblock \bibinfo{title}{Quantum Supremacy through the Quantum Approximate
  Optimization Algorithm}.
\newblock
\newblock
\showeprint{arXiv:1602.07674}


\bibitem[\protect\citeauthoryear{{Fu}, {Riesebos}, {Rol}, {van Straten}, {van
  Someren}, {Khammassi}, {Ashraf}, {Vermeulen}, {Newsum}, {Loh}, {de Sterke},
  {Vlothuizen}, {Schouten}, {Almudever}, {DiCarlo}, and {Bertels}}{{Fu}
  et~al\mbox{.}}{2019}]%
        {eQASM}
\bibfield{author}{\bibinfo{person}{X. {Fu}}, \bibinfo{person}{L. {Riesebos}},
  \bibinfo{person}{M.~A. {Rol}}, \bibinfo{person}{J. {van Straten}},
  \bibinfo{person}{J. {van Someren}}, \bibinfo{person}{N. {Khammassi}},
  \bibinfo{person}{I. {Ashraf}}, \bibinfo{person}{R.~F.~L. {Vermeulen}},
  \bibinfo{person}{V. {Newsum}}, \bibinfo{person}{K.~K.~L. {Loh}},
  \bibinfo{person}{J.~C. {de Sterke}}, \bibinfo{person}{W.~J. {Vlothuizen}},
  \bibinfo{person}{R.~N. {Schouten}}, \bibinfo{person}{C.~G. {Almudever}},
  \bibinfo{person}{L. {DiCarlo}}, {and} \bibinfo{person}{K. {Bertels}}.}
  \bibinfo{year}{2019}\natexlab{}.
\newblock \showarticletitle{eQASM: An Executable Quantum Instruction Set
  Architecture}. In \bibinfo{booktitle}{\emph{2019 IEEE International Symposium
  on High Performance Computer Architecture (HPCA)}}.
  \bibinfo{pages}{224--237}.
\newblock
\showISSN{2378-203X}
\urldef\tempurl%
\url{https://doi.org/10.1109/HPCA.2019.00040}
\showDOI{\tempurl}


\bibitem[\protect\citeauthoryear{Glaser, Boscain, Calarco, Koch,
  K{\"o}ckenberger, Kosloff, Kuprov, Luy, Schirmer, Schulte-Herbr{\"u}ggen,
  Sugny, and Wilhelm}{Glaser et~al\mbox{.}}{2015}]%
        {Glaser2015}
\bibfield{author}{\bibinfo{person}{Steffen~J. Glaser}, \bibinfo{person}{Ugo
  Boscain}, \bibinfo{person}{Tommaso Calarco}, \bibinfo{person}{Christiane~P.
  Koch}, \bibinfo{person}{Walter K{\"o}ckenberger}, \bibinfo{person}{Ronnie
  Kosloff}, \bibinfo{person}{Ilya Kuprov}, \bibinfo{person}{Burkhard Luy},
  \bibinfo{person}{Sophie Schirmer}, \bibinfo{person}{Thomas
  Schulte-Herbr{\"u}ggen}, \bibinfo{person}{Dominique Sugny}, {and}
  \bibinfo{person}{Frank~K. Wilhelm}.} \bibinfo{year}{2015}\natexlab{}.
\newblock \showarticletitle{Training Schr{\"o}dinger's cat: quantum optimal
  control}.
\newblock \bibinfo{journal}{\emph{The European Physical Journal D}}
  \bibinfo{volume}{69}, \bibinfo{number}{12} (\bibinfo{date}{17 Dec}
  \bibinfo{year}{2015}), \bibinfo{pages}{279}.
\newblock
\showISSN{1434-6079}
\urldef\tempurl%
\url{https://doi.org/10.1140/epjd/e2015-60464-1}
\showDOI{\tempurl}


\bibitem[\protect\citeauthoryear{Gokhale, Ding, Propson, and Winkler}{Gokhale
  et~al\mbox{.}}{2019}]%
        {GithubRepo}
\bibfield{author}{\bibinfo{person}{Pranav Gokhale}, \bibinfo{person}{Yongshan
  Ding}, \bibinfo{person}{Thomas Propson}, {and} \bibinfo{person}{Christopher
  Winkler}.} \bibinfo{year}{2019}\natexlab{}.
\newblock \bibinfo{title}{Code and Results: Partial Compilation of Variational
  Algorithms}.
\newblock
  \bibinfo{howpublished}{\url{https://github.com/EPiQC/PartialCompilation}}.
\newblock


\bibitem[\protect\citeauthoryear{Green, Lumsdaine, Ross, Selinger, and
  Valiron}{Green et~al\mbox{.}}{2013}]%
        {Quipper}
\bibfield{author}{\bibinfo{person}{Alexander~S. Green},
  \bibinfo{person}{Peter~LeFanu Lumsdaine}, \bibinfo{person}{Neil~J. Ross},
  \bibinfo{person}{Peter Selinger}, {and} \bibinfo{person}{Beno\^{\i}t
  Valiron}.} \bibinfo{year}{2013}\natexlab{}.
\newblock \showarticletitle{Quipper: A Scalable Quantum Programming Language}.
  In \bibinfo{booktitle}{\emph{Proceedings of the 34th ACM SIGPLAN Conference
  on Programming Language Design and Implementation}}
  \emph{(\bibinfo{series}{PLDI '13})}. \bibinfo{publisher}{ACM},
  \bibinfo{address}{New York, NY, USA}, \bibinfo{pages}{333--342}.
\newblock
\showISBNx{978-1-4503-2014-6}
\urldef\tempurl%
\url{https://doi.org/10.1145/2491956.2462177}
\showDOI{\tempurl}


\bibitem[\protect\citeauthoryear{Grover}{Grover}{1996}]%
        {Grover}
\bibfield{author}{\bibinfo{person}{Lov~K. Grover}.}
  \bibinfo{year}{1996}\natexlab{}.
\newblock \showarticletitle{A Fast Quantum Mechanical Algorithm for Database
  Search}. In \bibinfo{booktitle}{\emph{Annual ACM Symposium on Theory of
  Computing}}. \bibinfo{publisher}{ACM}, \bibinfo{pages}{212--219}.
\newblock


\bibitem[\protect\citeauthoryear{IBM}{IBM}{2019}]%
        {QiskitRoadmap}
\bibfield{author}{\bibinfo{person}{IBM}.} \bibinfo{year}{2019}\natexlab{}.
\newblock \bibinfo{title}{The Qiskit Roadmap 2019}.
\newblock
  \bibinfo{howpublished}{\url{https://github.com/Qiskit/qiskit/blob/master/docs/development_strategy.rst}}.
\newblock


\bibitem[\protect\citeauthoryear{Javadi-Abhari, Faruque, Dousti, Svec, Catu,
  Chakrabati, Chiang, Vanderwilt, Black, Chong, Martonosi, Suchara, Brown,
  Pedram, and Brun}{Javadi-Abhari et~al\mbox{.}}{2012}]%
        {Scaffold}
\bibfield{author}{\bibinfo{person}{Ali Javadi-Abhari}, \bibinfo{person}{Arvin
  Faruque}, \bibinfo{person}{Mohammad~Javad Dousti}, \bibinfo{person}{Lukas
  Svec}, \bibinfo{person}{Oana Catu}, \bibinfo{person}{Amlan Chakrabati},
  \bibinfo{person}{Chen-Fu Chiang}, \bibinfo{person}{Seth Vanderwilt},
  \bibinfo{person}{John Black}, \bibinfo{person}{Fred Chong},
  \bibinfo{person}{Margaret Martonosi}, \bibinfo{person}{Martin Suchara},
  \bibinfo{person}{Ken Brown}, \bibinfo{person}{Massoud Pedram}, {and}
  \bibinfo{person}{Todd Brun}.} \bibinfo{year}{2012}\natexlab{}.
\newblock \bibinfo{title}{Scaffold: Quantum Programming Language}.
\newblock
\newblock


\bibitem[\protect\citeauthoryear{Javadi-Abhari, Patil, Kudrow, Heckey, Lvov,
  Chong, and Martonosi}{Javadi-Abhari et~al\mbox{.}}{2014}]%
        {ScaffCC}
\bibfield{author}{\bibinfo{person}{Ali Javadi-Abhari}, \bibinfo{person}{Shruti
  Patil}, \bibinfo{person}{Daniel Kudrow}, \bibinfo{person}{Jeff Heckey},
  \bibinfo{person}{Alexey Lvov}, \bibinfo{person}{Frederic~T. Chong}, {and}
  \bibinfo{person}{Margaret Martonosi}.} \bibinfo{year}{2014}\natexlab{}.
\newblock \showarticletitle{ScaffCC: A Framework for Compilation and Analysis
  of Quantum Computing Programs}. In \bibinfo{booktitle}{\emph{Proceedings of
  the 11th ACM Conference on Computing Frontiers}} \emph{(\bibinfo{series}{CF
  '14})}. \bibinfo{publisher}{ACM}, \bibinfo{address}{New York, NY, USA},
  Article \bibinfo{articleno}{1}, \bibinfo{numpages}{10}~pages.
\newblock
\showISBNx{978-1-4503-2870-8}
\urldef\tempurl%
\url{https://doi.org/10.1145/2597917.2597939}
\showDOI{\tempurl}


\bibitem[\protect\citeauthoryear{Kandala, Mezzacapo, Temme, Takita, Brink,
  Chow, and Gambetta}{Kandala et~al\mbox{.}}{2017}]%
        {Kandala}
\bibfield{author}{\bibinfo{person}{Abhinav Kandala}, \bibinfo{person}{Antonio
  Mezzacapo}, \bibinfo{person}{Kristan Temme}, \bibinfo{person}{Maika Takita},
  \bibinfo{person}{Markus Brink}, \bibinfo{person}{Jerry~M. Chow}, {and}
  \bibinfo{person}{Jay~M. Gambetta}.} \bibinfo{year}{2017}\natexlab{}.
\newblock \showarticletitle{Hardware-efficient variational quantum eigensolver
  for small molecules and quantum magnets}.
\newblock \bibinfo{journal}{\emph{Nature}}  \bibinfo{volume}{549}
  (\bibinfo{date}{13 Sep} \bibinfo{year}{2017}), \bibinfo{pages}{242 EP --}.
\newblock
\urldef\tempurl%
\url{https://doi.org/10.1038/nature23879}
\showURL{%
\tempurl}


\bibitem[\protect\citeauthoryear{Khaneja, Reiss, Kehlet,
  Schulte-Herbr{\"u}ggen, and Glaser}{Khaneja et~al\mbox{.}}{2005}]%
        {GRAPE}
\bibfield{author}{\bibinfo{person}{Navin Khaneja}, \bibinfo{person}{Timo
  Reiss}, \bibinfo{person}{Cindie Kehlet}, \bibinfo{person}{Thomas
  Schulte-Herbr{\"u}ggen}, {and} \bibinfo{person}{Steffen~J. Glaser}.}
  \bibinfo{year}{2005}\natexlab{}.
\newblock \showarticletitle{Optimal control of coupled spin dynamics: design of
  NMR pulse sequences by gradient ascent algorithms}.
\newblock \bibinfo{journal}{\emph{Journal of Magnetic Resonance}}
  \bibinfo{volume}{172}, \bibinfo{number}{2} (\bibinfo{year}{2005}),
  \bibinfo{pages}{296 -- 305}.
\newblock
\showISSN{1090-7807}
\urldef\tempurl%
\url{https://doi.org/10.1016/j.jmr.2004.11.004}
\showDOI{\tempurl}


\bibitem[\protect\citeauthoryear{Kudrow, Bier, Deng, Franklin, Tomita, Brown,
  and Chong}{Kudrow et~al\mbox{.}}{2013}]%
        {QuantumRotations}
\bibfield{author}{\bibinfo{person}{Daniel Kudrow}, \bibinfo{person}{Kenneth
  Bier}, \bibinfo{person}{Zhaoxia Deng}, \bibinfo{person}{Diana Franklin},
  \bibinfo{person}{Yu Tomita}, \bibinfo{person}{Kenneth~R. Brown}, {and}
  \bibinfo{person}{Frederic~T. Chong}.} \bibinfo{year}{2013}\natexlab{}.
\newblock \showarticletitle{Quantum Rotations: A Case Study in Static and
  Dynamic Machine-code Generation for Quantum Computers}. In
  \bibinfo{booktitle}{\emph{Proceedings of the 40th Annual International
  Symposium on Computer Architecture}} \emph{(\bibinfo{series}{ISCA '13})}.
  \bibinfo{publisher}{ACM}, \bibinfo{address}{New York, NY, USA},
  \bibinfo{pages}{166--176}.
\newblock
\showISBNx{978-1-4503-2079-5}
\urldef\tempurl%
\url{https://doi.org/10.1145/2485922.2485937}
\showDOI{\tempurl}


\bibitem[\protect\citeauthoryear{Leung, Abdelhafez, Koch, and Schuster}{Leung
  et~al\mbox{.}}{2017}]%
        {NelsonPaper}
\bibfield{author}{\bibinfo{person}{Nelson Leung}, \bibinfo{person}{Mohamed
  Abdelhafez}, \bibinfo{person}{Jens Koch}, {and} \bibinfo{person}{David
  Schuster}.} \bibinfo{year}{2017}\natexlab{}.
\newblock \showarticletitle{Speedup for quantum optimal control from automatic
  differentiation based on graphics processing units}.
\newblock \bibinfo{journal}{\emph{Phys. Rev. A}}  \bibinfo{volume}{95}
  (\bibinfo{date}{Apr} \bibinfo{year}{2017}), \bibinfo{pages}{042318}.
\newblock
Issue 4.
\urldef\tempurl%
\url{https://doi.org/10.1103/PhysRevA.95.042318}
\showDOI{\tempurl}


\bibitem[\protect\citeauthoryear{Lloyd}{Lloyd}{1996}]%
        {lloyd1996universal}
\bibfield{author}{\bibinfo{person}{Seth Lloyd}.}
  \bibinfo{year}{1996}\natexlab{}.
\newblock \showarticletitle{Universal quantum simulators}.
\newblock \bibinfo{journal}{\emph{Science}} (\bibinfo{year}{1996}),
  \bibinfo{pages}{1073--1078}.
\newblock


\bibitem[\protect\citeauthoryear{Lloyd}{Lloyd}{2018}]%
        {QAOA_Universality}
\bibfield{author}{\bibinfo{person}{Seth Lloyd}.}
  \bibinfo{year}{2018}\natexlab{}.
\newblock \bibinfo{title}{Quantum approximate optimization is computationally
  universal}.
\newblock
\newblock
\showeprint{arXiv:1812.11075}


\bibitem[\protect\citeauthoryear{Lloyd and Maity}{Lloyd and Maity}{2019}]%
        {LloydQOC}
\bibfield{author}{\bibinfo{person}{Seth Lloyd} {and} \bibinfo{person}{Reevu
  Maity}.} \bibinfo{year}{2019}\natexlab{}.
\newblock \bibinfo{title}{Efficient implementation of unitary transformations}.
\newblock
\newblock
\showeprint{arXiv:1901.03431}


\bibitem[\protect\citeauthoryear{McArdle, Endo, Aspuru-Guzik, Benjamin, and
  Yuan}{McArdle et~al\mbox{.}}{2018}]%
        {Quantum_Computational_Chemistry}
\bibfield{author}{\bibinfo{person}{Sam McArdle}, \bibinfo{person}{Suguru Endo},
  \bibinfo{person}{Alan Aspuru-Guzik}, \bibinfo{person}{Simon Benjamin}, {and}
  \bibinfo{person}{Xiao Yuan}.} \bibinfo{year}{2018}\natexlab{}.
\newblock \bibinfo{title}{Quantum computational chemistry}.
\newblock
\newblock
\showeprint{arXiv:1808.10402}


\bibitem[\protect\citeauthoryear{McClean, Romero, Babbush, and
  Aspuru-Guzik}{McClean et~al\mbox{.}}{2016}]%
        {McClean_2016}
\bibfield{author}{\bibinfo{person}{Jarrod~R McClean}, \bibinfo{person}{Jonathan
  Romero}, \bibinfo{person}{Ryan Babbush}, {and} \bibinfo{person}{Al{\'{a}}n
  Aspuru-Guzik}.} \bibinfo{year}{2016}\natexlab{}.
\newblock \showarticletitle{The theory of variational hybrid quantum-classical
  algorithms}.
\newblock \bibinfo{journal}{\emph{New Journal of Physics}}
  \bibinfo{volume}{18}, \bibinfo{number}{2} (\bibinfo{date}{feb}
  \bibinfo{year}{2016}), \bibinfo{pages}{023023}.
\newblock
\urldef\tempurl%
\url{https://doi.org/10.1088/1367-2630/18/2/023023}
\showDOI{\tempurl}


\bibitem[\protect\citeauthoryear{McKay, Alexander, Bello, Biercuk, Bishop,
  Chen, Chow, C{\'o}rcoles, Egger, Filipp, Gomez, Hush, Javadi-Abhari, Moreda,
  Nation, Paulovicks, Winston, Wood, Wootton, and Gambetta}{McKay
  et~al\mbox{.}}{2018}]%
        {OpenPulse}
\bibfield{author}{\bibinfo{person}{David~C. McKay}, \bibinfo{person}{Thomas
  Alexander}, \bibinfo{person}{Luciano Bello}, \bibinfo{person}{Michael~J.
  Biercuk}, \bibinfo{person}{Lev Bishop}, \bibinfo{person}{Jiayin Chen},
  \bibinfo{person}{Jerry~M. Chow}, \bibinfo{person}{Antonio~D. C{\'o}rcoles},
  \bibinfo{person}{Daniel Egger}, \bibinfo{person}{Stefan Filipp},
  \bibinfo{person}{Juan Gomez}, \bibinfo{person}{Michael Hush},
  \bibinfo{person}{Ali Javadi-Abhari}, \bibinfo{person}{Diego Moreda},
  \bibinfo{person}{Paul Nation}, \bibinfo{person}{Brent Paulovicks},
  \bibinfo{person}{Erick Winston}, \bibinfo{person}{Christopher~J. Wood},
  \bibinfo{person}{James Wootton}, {and} \bibinfo{person}{Jay~M. Gambetta}.}
  \bibinfo{year}{2018}\natexlab{}.
\newblock \bibinfo{title}{Qiskit Backend Specifications for OpenQASM and
  OpenPulse Experiments}.
\newblock
\newblock
\showeprint{arXiv:1809.03452}


\bibitem[\protect\citeauthoryear{McKay, Wood, Sheldon, Chow, and
  Gambetta}{McKay et~al\mbox{.}}{2017}]%
        {PhysRevA.96.022330}
\bibfield{author}{\bibinfo{person}{David~C. McKay},
  \bibinfo{person}{Christopher~J. Wood}, \bibinfo{person}{Sarah Sheldon},
  \bibinfo{person}{Jerry~M. Chow}, {and} \bibinfo{person}{Jay~M. Gambetta}.}
  \bibinfo{year}{2017}\natexlab{}.
\newblock \showarticletitle{Efficient $Z$ gates for quantum computing}.
\newblock \bibinfo{journal}{\emph{Phys. Rev. A}}  \bibinfo{volume}{96}
  (\bibinfo{date}{Aug} \bibinfo{year}{2017}), \bibinfo{pages}{022330}.
\newblock
Issue 2.
\urldef\tempurl%
\url{https://doi.org/10.1103/PhysRevA.96.022330}
\showDOI{\tempurl}


\bibitem[\protect\citeauthoryear{Moll, Barkoutsos, Bishop, Chow, Cross, Egger,
  Filipp, Fuhrer, Gambetta, Ganzhorn, Kandala, Mezzacapo, M{\"u}ller, Riess,
  Salis, Smolin, Tavernelli, and Temme}{Moll et~al\mbox{.}}{2018}]%
        {Moll_2018}
\bibfield{author}{\bibinfo{person}{Nikolaj Moll}, \bibinfo{person}{Panagiotis
  Barkoutsos}, \bibinfo{person}{Lev~S Bishop}, \bibinfo{person}{Jerry~M Chow},
  \bibinfo{person}{Andrew Cross}, \bibinfo{person}{Daniel~J Egger},
  \bibinfo{person}{Stefan Filipp}, \bibinfo{person}{Andreas Fuhrer},
  \bibinfo{person}{Jay~M Gambetta}, \bibinfo{person}{Marc Ganzhorn},
  \bibinfo{person}{Abhinav Kandala}, \bibinfo{person}{Antonio Mezzacapo},
  \bibinfo{person}{Peter M{\"u}ller}, \bibinfo{person}{Walter Riess},
  \bibinfo{person}{Gian Salis}, \bibinfo{person}{John Smolin},
  \bibinfo{person}{Ivano Tavernelli}, {and} \bibinfo{person}{Kristan Temme}.}
  \bibinfo{year}{2018}\natexlab{}.
\newblock \showarticletitle{Quantum optimization using variational algorithms
  on near-term quantum devices}.
\newblock \bibinfo{journal}{\emph{Quantum Science and Technology}}
  \bibinfo{volume}{3}, \bibinfo{number}{3} (\bibinfo{date}{jun}
  \bibinfo{year}{2018}), \bibinfo{pages}{030503}.
\newblock
\urldef\tempurl%
\url{https://doi.org/10.1088/2058-9565/aab822}
\showDOI{\tempurl}


\bibitem[\protect\citeauthoryear{Nam, Chen, Pisenti, Wright, Delaney, Maslov,
  Brown, Allen, Amini, Apisdorf, Beck, Blinov, Chaplin, Chmielewski, Collins,
  Debnath, Ducore, Hudek, Keesan, Kreikemeier, Mizrahi, Solomon, Williams,
  Wong-Campos, Monroe, and Kim}{Nam et~al\mbox{.}}{2019}]%
        {WaterVQE}
\bibfield{author}{\bibinfo{person}{Yunseong Nam}, \bibinfo{person}{Jwo-Sy
  Chen}, \bibinfo{person}{Neal~C. Pisenti}, \bibinfo{person}{Kenneth Wright},
  \bibinfo{person}{Conor Delaney}, \bibinfo{person}{Dmitri Maslov},
  \bibinfo{person}{Kenneth~R. Brown}, \bibinfo{person}{Stewart Allen},
  \bibinfo{person}{Jason~M. Amini}, \bibinfo{person}{Joel Apisdorf},
  \bibinfo{person}{Kristin~M. Beck}, \bibinfo{person}{Aleksey Blinov},
  \bibinfo{person}{Vandiver Chaplin}, \bibinfo{person}{Mika Chmielewski},
  \bibinfo{person}{Coleman Collins}, \bibinfo{person}{Shantanu Debnath},
  \bibinfo{person}{Andrew~M. Ducore}, \bibinfo{person}{Kai~M. Hudek},
  \bibinfo{person}{Matthew Keesan}, \bibinfo{person}{Sarah~M. Kreikemeier},
  \bibinfo{person}{Jonathan Mizrahi}, \bibinfo{person}{Phil Solomon},
  \bibinfo{person}{Mike Williams}, \bibinfo{person}{Jaime~David Wong-Campos},
  \bibinfo{person}{Christopher Monroe}, {and} \bibinfo{person}{Jungsang Kim}.}
  \bibinfo{year}{2019}\natexlab{}.
\newblock \bibinfo{title}{Ground-state energy estimation of the water molecule
  on a trapped ion quantum computer}.
\newblock
\newblock
\showeprint{arXiv:1902.10171}


\bibitem[\protect\citeauthoryear{Nielsen and Chuang}{Nielsen and
  Chuang}{2011}]%
        {Nielsen}
\bibfield{author}{\bibinfo{person}{Michael~A. Nielsen} {and}
  \bibinfo{person}{Isaac~L. Chuang}.} \bibinfo{year}{2011}\natexlab{}.
\newblock \bibinfo{booktitle}{\emph{Quantum Computation and Quantum
  Information: 10th Anniversary Edition} (\bibinfo{edition}{10th} ed.)}.
\newblock \bibinfo{publisher}{Cambridge University Press},
  \bibinfo{address}{New York, NY, USA}.
\newblock
\showISBNx{1107002176, 9781107002173}


\bibitem[\protect\citeauthoryear{O'Gorman and Campbell}{O'Gorman and
  Campbell}{2017}]%
        {MagicStateOverhead}
\bibfield{author}{\bibinfo{person}{Joe O'Gorman} {and} \bibinfo{person}{Earl~T.
  Campbell}.} \bibinfo{year}{2017}\natexlab{}.
\newblock \showarticletitle{Quantum computation with realistic magic-state
  factories}.
\newblock \bibinfo{journal}{\emph{Phys. Rev. A}}  \bibinfo{volume}{95}
  (\bibinfo{date}{Mar} \bibinfo{year}{2017}), \bibinfo{pages}{032338}.
\newblock
Issue 3.
\urldef\tempurl%
\url{https://doi.org/10.1103/PhysRevA.95.032338}
\showDOI{\tempurl}


\bibitem[\protect\citeauthoryear{Otterbach, Manenti, Alidoust, Bestwick, Block,
  Bloom, Caldwell, Didier, Fried, Hong, Karalekas, Osborn, Papageorge,
  Peterson, Prawiroatmodjo, Rubin, Ryan, Scarabelli, Scheer, Sete, Sivarajah,
  Smith, Staley, Tezak, Zeng, Hudson, Johnson, Reagor, da~Silva, and
  Rigetti}{Otterbach et~al\mbox{.}}{2017}]%
        {Rigetti_QAOA_Experiment}
\bibfield{author}{\bibinfo{person}{J.~S. Otterbach}, \bibinfo{person}{R.
  Manenti}, \bibinfo{person}{N. Alidoust}, \bibinfo{person}{A. Bestwick},
  \bibinfo{person}{M. Block}, \bibinfo{person}{B. Bloom}, \bibinfo{person}{S.
  Caldwell}, \bibinfo{person}{N. Didier}, \bibinfo{person}{E.~Schuyler Fried},
  \bibinfo{person}{S. Hong}, \bibinfo{person}{P. Karalekas},
  \bibinfo{person}{C.~B. Osborn}, \bibinfo{person}{A. Papageorge},
  \bibinfo{person}{E.~C. Peterson}, \bibinfo{person}{G. Prawiroatmodjo},
  \bibinfo{person}{N. Rubin}, \bibinfo{person}{Colm~A. Ryan},
  \bibinfo{person}{D. Scarabelli}, \bibinfo{person}{M. Scheer},
  \bibinfo{person}{E.~A. Sete}, \bibinfo{person}{P. Sivarajah},
  \bibinfo{person}{Robert~S. Smith}, \bibinfo{person}{A. Staley},
  \bibinfo{person}{N. Tezak}, \bibinfo{person}{W.~J. Zeng}, \bibinfo{person}{A.
  Hudson}, \bibinfo{person}{Blake~R. Johnson}, \bibinfo{person}{M. Reagor},
  \bibinfo{person}{M.~P. da Silva}, {and} \bibinfo{person}{C. Rigetti}.}
  \bibinfo{year}{2017}\natexlab{}.
\newblock \bibinfo{title}{Unsupervised Machine Learning on a Hybrid Quantum
  Computer}.
\newblock
\newblock
\showeprint{arXiv:1712.05771}


\bibitem[\protect\citeauthoryear{Paulsen and Trautwein}{Paulsen and
  Trautwein}{2004}]%
        {1206.2247}
\bibfield{author}{\bibinfo{person}{Hauke Paulsen} {and}
  \bibinfo{person}{Alfred~X Trautwein}.} \bibinfo{year}{2004}\natexlab{}.
\newblock \showarticletitle{Density functional theory calculations for spin
  crossover complexes}.
\newblock  (\bibinfo{year}{2004}), \bibinfo{pages}{197--219}.
\newblock


\bibitem[\protect\citeauthoryear{Peruzzo, McClean, Shadbolt, Yung, Zhou, Love,
  Aspuru-Guzik, and O'Brien}{Peruzzo et~al\mbox{.}}{2014}]%
        {VQE}
\bibfield{author}{\bibinfo{person}{Alberto Peruzzo}, \bibinfo{person}{Jarrod
  McClean}, \bibinfo{person}{Peter Shadbolt}, \bibinfo{person}{Man-Hong Yung},
  \bibinfo{person}{Xiao-Qi Zhou}, \bibinfo{person}{Peter~J. Love},
  \bibinfo{person}{Al{\'a}n Aspuru-Guzik}, {and} \bibinfo{person}{Jeremy~L.
  O'Brien}.} \bibinfo{year}{2014}\natexlab{}.
\newblock \showarticletitle{A variational eigenvalue solver on a photonic
  quantum processor}.
\newblock \bibinfo{journal}{\emph{Nature Communications}}  \bibinfo{volume}{5}
  (\bibinfo{date}{23 Jul} \bibinfo{year}{2014}), \bibinfo{pages}{4213 EP --}.
\newblock
\urldef\tempurl%
\url{https://doi.org/10.1038/ncomms5213}
\showURL{%
\tempurl}
\newblock
\shownote{Article.}


\bibitem[\protect\citeauthoryear{Preskill}{Preskill}{2018}]%
        {NISQ}
\bibfield{author}{\bibinfo{person}{John Preskill}.}
  \bibinfo{year}{2018}\natexlab{}.
\newblock \showarticletitle{Quantum {C}omputing in the {NISQ} era and beyond}.
\newblock \bibinfo{journal}{\emph{{Quantum}}}  \bibinfo{volume}{2}
  (\bibinfo{date}{Aug.} \bibinfo{year}{2018}), \bibinfo{pages}{79}.
\newblock
\showISSN{2521-327X}
\urldef\tempurl%
\url{https://doi.org/10.22331/q-2018-08-06-79}
\showDOI{\tempurl}


\bibitem[\protect\citeauthoryear{Romero, Babbush, McClean, Hempel, Love, and
  Aspuru-Guzik}{Romero et~al\mbox{.}}{2018}]%
        {UCCSD_Strategies}
\bibfield{author}{\bibinfo{person}{Jonathan Romero}, \bibinfo{person}{Ryan
  Babbush}, \bibinfo{person}{Jarrod~R McClean}, \bibinfo{person}{Cornelius
  Hempel}, \bibinfo{person}{Peter~J Love}, {and} \bibinfo{person}{Al{\'{a}}n
  Aspuru-Guzik}.} \bibinfo{year}{2018}\natexlab{}.
\newblock \showarticletitle{Strategies for quantum computing molecular energies
  using the unitary coupled cluster ansatz}.
\newblock \bibinfo{journal}{\emph{Quantum Science and Technology}}
  \bibinfo{volume}{4}, \bibinfo{number}{1} (\bibinfo{date}{oct}
  \bibinfo{year}{2018}), \bibinfo{pages}{014008}.
\newblock
\urldef\tempurl%
\url{https://doi.org/10.1088/2058-9565/aad3e4}
\showDOI{\tempurl}


\bibitem[\protect\citeauthoryear{Shi, Leung, Gokhale, Rossi, Schuster,
  Hoffmann, and Chong}{Shi et~al\mbox{.}}{2019}]%
        {YunongPaper}
\bibfield{author}{\bibinfo{person}{Yunong Shi}, \bibinfo{person}{Nelson Leung},
  \bibinfo{person}{Pranav Gokhale}, \bibinfo{person}{Zane Rossi},
  \bibinfo{person}{David~I Schuster}, \bibinfo{person}{Henry Hoffmann}, {and}
  \bibinfo{person}{Frederic~T Chong}.} \bibinfo{year}{2019}\natexlab{}.
\newblock \showarticletitle{Optimized Compilation of Aggregated Instructions
  for Realistic Quantum Computers}. In \bibinfo{booktitle}{\emph{Proceedings of
  the Twenty-Fourth International Conference on Architectural Support for
  Programming Languages and Operating Systems}}. ACM,
  \bibinfo{pages}{1031--1044}.
\newblock


\bibitem[\protect\citeauthoryear{Shor}{Shor}{1997}]%
        {Shor}
\bibfield{author}{\bibinfo{person}{Peter~W. Shor}.}
  \bibinfo{year}{1997}\natexlab{}.
\newblock \showarticletitle{Polynomial-Time Algorithms for Prime Factorization
  and Discrete Logarithms on a Quantum Computer}.
\newblock \bibinfo{journal}{\emph{SIAM J. Comput.}} \bibinfo{volume}{26},
  \bibinfo{number}{5} (\bibinfo{date}{Oct.} \bibinfo{year}{1997}),
  \bibinfo{pages}{1484--1509}.
\newblock
\showISSN{0097-5397}
\urldef\tempurl%
\url{https://doi.org/10.1137/S0097539795293172}
\showDOI{\tempurl}


\bibitem[\protect\citeauthoryear{Smith, Curtis, and Zeng}{Smith
  et~al\mbox{.}}{2016}]%
        {Quil}
\bibfield{author}{\bibinfo{person}{Robert~S. Smith},
  \bibinfo{person}{Michael~J. Curtis}, {and} \bibinfo{person}{William~J.
  Zeng}.} \bibinfo{year}{2016}\natexlab{}.
\newblock \showarticletitle{A Practical Quantum Instruction Set Architecture}.
\newblock \bibinfo{journal}{\emph{CoRR}}  \bibinfo{volume}{abs/1608.03355}
  (\bibinfo{year}{2016}).
\newblock


\bibitem[\protect\citeauthoryear{Steiger, H{\"{a}}ner, and Troyer}{Steiger
  et~al\mbox{.}}{2018}]%
        {ProjectQ}
\bibfield{author}{\bibinfo{person}{Damian~S. Steiger}, \bibinfo{person}{Thomas
  H{\"{a}}ner}, {and} \bibinfo{person}{Matthias Troyer}.}
  \bibinfo{year}{2018}\natexlab{}.
\newblock \showarticletitle{Project{Q}: an open source software framework for
  quantum computing}.
\newblock \bibinfo{journal}{\emph{{Quantum}}}  \bibinfo{volume}{2}
  (\bibinfo{date}{Jan.} \bibinfo{year}{2018}), \bibinfo{pages}{49}.
\newblock
\showISSN{2521-327X}
\urldef\tempurl%
\url{https://doi.org/10.22331/q-2018-01-31-49}
\showDOI{\tempurl}


\bibitem[\protect\citeauthoryear{Suchara, Faruque, Lai, Paz, Chong, and
  Kubiatowicz}{Suchara et~al\mbox{.}}{2013}]%
        {ErrorCorrectionOverhead}
\bibfield{author}{\bibinfo{person}{Martin Suchara}, \bibinfo{person}{Arvin
  Faruque}, \bibinfo{person}{Ching-Yi Lai}, \bibinfo{person}{Gerardo Paz},
  \bibinfo{person}{Frederic~T. Chong}, {and} \bibinfo{person}{John
  Kubiatowicz}.} \bibinfo{year}{2013}\natexlab{}.
\newblock \showarticletitle{Comparing the Overhead of Topological and
  Concatenated Quantum Error Correction}.
\newblock \bibinfo{howpublished}{Part of the work was in Proceedings of IEEE
  International Conference on Computer Design (ICCD) 2013}.
\newblock \bibinfo{journal}{\emph{arXiv preprint arXiv:1312.2316}}
  (\bibinfo{year}{2013}).
\newblock
\urldef\tempurl%
\url{https://doi.org/10.1109/ICCD.2013.6657074}
\showDOI{\tempurl}
\showeprint{arXiv:1312.2316}


\bibitem[\protect\citeauthoryear{Sun, Berkelbach, Blunt, Booth, Guo, Li, Liu,
  McClain, Sayfutyarova, Sharma, Wouters, and Chan}{Sun et~al\mbox{.}}{2017}]%
        {PySCF}
\bibfield{author}{\bibinfo{person}{Qiming Sun}, \bibinfo{person}{Timothy~C.
  Berkelbach}, \bibinfo{person}{Nick~S. Blunt}, \bibinfo{person}{George~H.
  Booth}, \bibinfo{person}{Sheng Guo}, \bibinfo{person}{Zhendong Li},
  \bibinfo{person}{Junzi Liu}, \bibinfo{person}{James~D. McClain},
  \bibinfo{person}{Elvira~R. Sayfutyarova}, \bibinfo{person}{Sandeep Sharma},
  \bibinfo{person}{Sebastian Wouters}, {and} \bibinfo{person}{Garnet Kin-Lic
  Chan}.} \bibinfo{year}{2017}\natexlab{}.
\newblock \bibinfo{title}{PySCF: the Python-based simulations of chemistry
  framework}.
\newblock , \bibinfo{numpages}{e1340}~pages.
\newblock
\urldef\tempurl%
\url{https://doi.org/10.1002/wcms.1340}
\showDOI{\tempurl}
\showeprint{https://onlinelibrary.wiley.com/doi/pdf/10.1002/wcms.1340}


\bibitem[\protect\citeauthoryear{Svore, Geller, Troyer, Azariah, Granade, Heim,
  Kliuchnikov, Mykhailova, Paz, and Roetteler}{Svore et~al\mbox{.}}{2018}]%
        {QSharp}
\bibfield{author}{\bibinfo{person}{Krysta Svore}, \bibinfo{person}{Alan
  Geller}, \bibinfo{person}{Matthias Troyer}, \bibinfo{person}{John Azariah},
  \bibinfo{person}{Christopher Granade}, \bibinfo{person}{Bettina Heim},
  \bibinfo{person}{Vadym Kliuchnikov}, \bibinfo{person}{Mariia Mykhailova},
  \bibinfo{person}{Andres Paz}, {and} \bibinfo{person}{Martin Roetteler}.}
  \bibinfo{year}{2018}\natexlab{}.
\newblock \showarticletitle{Q\#: Enabling Scalable Quantum Computing and
  Development with a High-level DSL}. In \bibinfo{booktitle}{\emph{Proceedings
  of the Real World Domain Specific Languages Workshop 2018}}
  \emph{(\bibinfo{series}{RWDSL2018})}. \bibinfo{publisher}{ACM},
  \bibinfo{address}{New York, NY, USA}, Article \bibinfo{articleno}{7},
  \bibinfo{numpages}{10}~pages.
\newblock
\showISBNx{978-1-4503-6355-6}
\urldef\tempurl%
\url{https://doi.org/10.1145/3183895.3183901}
\showDOI{\tempurl}


\bibitem[\protect\citeauthoryear{Thornton, Hutter, Hoos, and
  Leyton-Brown}{Thornton et~al\mbox{.}}{2013}]%
        {thornton2013auto}
\bibfield{author}{\bibinfo{person}{Chris Thornton}, \bibinfo{person}{Frank
  Hutter}, \bibinfo{person}{Holger~H Hoos}, {and} \bibinfo{person}{Kevin
  Leyton-Brown}.} \bibinfo{year}{2013}\natexlab{}.
\newblock \showarticletitle{Auto-WEKA: Combined selection and hyperparameter
  optimization of classification algorithms}. In
  \bibinfo{booktitle}{\emph{Proceedings of the 19th ACM SIGKDD international
  conference on Knowledge discovery and data mining}}. ACM,
  \bibinfo{pages}{847--855}.
\newblock


\bibitem[\protect\citeauthoryear{Wang, Higgott, and Brierley}{Wang
  et~al\mbox{.}}{2018}]%
        {wang2018generalised}
\bibfield{author}{\bibinfo{person}{Daochen Wang}, \bibinfo{person}{Oscar
  Higgott}, {and} \bibinfo{person}{Stephen Brierley}.}
  \bibinfo{year}{2018}\natexlab{}.
\newblock \showarticletitle{A Generalised Variational Quantum Eigensolver}.
\newblock \bibinfo{journal}{\emph{arXiv preprint arXiv:1802.00171}}
  (\bibinfo{year}{2018}).
\newblock


\bibitem[\protect\citeauthoryear{Wecker and Svore}{Wecker and Svore}{2014}]%
        {LIQUI}
\bibfield{author}{\bibinfo{person}{Dave Wecker} {and}
  \bibinfo{person}{Krysta~M. Svore}.} \bibinfo{year}{2014}\natexlab{}.
\newblock \bibinfo{title}{LIQUi|>: A Software Design Architecture and
  Domain-Specific Language for Quantum Computing}.
\newblock
\newblock
\showeprint{arXiv:1402.4467}


\end{thebibliography}
